\documentclass[11pt]{article}

\usepackage{cite}
\usepackage{url}
\usepackage{microtype}

\usepackage{amsmath,amssymb,amsthm,latexsym, color,graphicx, xspace}

\def\reals{\mathbb{R}}

\def\comp{{\mathbb C}}

\def\ttt{{\bf t}}

\let\eps\varepsilon

\def\etal{\textit{et~al.}}

\providecommand{\remove}[1]{}

\newtheorem{theorem}{Theorem}
\newtheorem{lemma}[theorem]{Lemma}
\newtheorem{claim}[theorem]{Claim}
\newtheorem{proposition}[theorem]{Proposition}

\newtheorem{corollary}[theorem]{Corollary}
\newtheorem{conjecture}[theorem]{Conjecture}

\theoremstyle{remark}

\newcommand{\ignore}[1]{}

\begin{document}

\begin{titlepage}

\title{Incidences in Three Dimensions and Distinct Distances in the
Plane\thanks{%
  Work by Micha Sharir has been supported 
  by NSF Grants CCF-05-14079 and CCF-08-30272, 
  by grant 2006/194 from the U.S.-Israeli Binational Science Foundation,
  by grants 155/05 and 338/09 from the Israel Science Fund, 
  and by the Hermann Minkowski--MINERVA Center for Geometry at Tel Aviv
  University.
  } }

\author{Gy\"orgy Elekes\thanks{%
Department of Computer Science, E\"otv\"os University, Budapest.} 
\and
Micha Sharir\thanks{%
School of Computer Science, Tel Aviv University,
Tel Aviv 69978 Israel and Courant Institute of Mathematical Sciences,
New York University, New York, NY 10012, USA;
\textsl{michas@post.tau.ac.il}.}
}

\maketitle

\begin{abstract}
  We first describe a reduction from the problem of lower-bounding the
  number of distinct distances determined by a set $S$ of $s$ points 
  in the plane to 
  an incidence problem between points and a certain class of helices
  (or parabolas) in three dimensions. We offer conjectures involving 
  the new setup, but are still unable to fully resolve them. 

  Instead, we adapt the recent new algebraic analysis technique of Guth and
  Katz~\cite{GK}, as further developed by Elekes et al.~\cite{EKS}, 
  to obtain sharp bounds on the number of incidences between these
  helices or parabolas and points in $\reals^3$. 
  Applying these bounds, we obtain, among several other results,
  the upper bound $O(s^3)$ on the number of rotations (rigid motions) 
  which map (at least) three points of $S$ to three other points of
  $S$. In fact, we show that the number of such rotations which
  map at least $k\ge 3$ points of $S$ to $k$ other points of $S$ is
  close to $O(s^3/k^{12/7})$. 

  One of our unresolved conjectures is that this number is
  $O(s^3/k^2)$, for $k\ge 2$. If true, it would imply the lower bound
  $\Omega(s/\log s)$ on the number of distinct distances in the plane.
\end{abstract}

\thispagestyle{empty}
\end{titlepage}

\section{The infrastructure} 
\label{sec:infra}

The motivation for the study reported in this paper comes from the
celebrated and long-standing problem, originally posed by Erd{\H o}s 
\cite{Er} in 1946, of obtaining a sharp lower bound for the number of
distinct distances guaranteed to exist in any set $S$ of $s$ points 
in the plane. Erd{\H o}s has shown that a section of the integer
lattice determines only $O(s/\sqrt{\log s})$ distinct distances, and
conjectured this to be a lower bound for any planar point set. 
In spite of steady progress on this problem, reviewed next,
Erd{\H o}s's conjecture is still open.

L. Moser \cite{Mo}, Chung \cite{Chu},
and Chung~\etal~\cite{ChSzT} proved that the number of distinct 
distances determined by $s$ points in the plane is $\Omega(s^{2/3})$, 
$\Omega(s^{5/7})$, and $\Omega(s^{4/5}/{\rm polylog}(s))$, 
respectively. Sz\'ekely \cite{Sz} managed to get rid
of the polylogarithmic factor, while Solymosi and T\'oth
\cite{SoTo} improved this bound to $\Omega(s^{6/7})$. This
was a real breakthrough. Their analysis was subsequently refined by
Tardos \cite{Ta} and then by Katz and Tardos \cite{KT}, who obtained the
current record of $\Omega(s^{(48-14e)/(55-16e)-\eps})$, for any
$\eps>0$, which is $\Omega(s^{0.8641})$. 

In this paper we transform the problem of distinct distances in the
plane to an incidence problem between points and a certain kind of
curves (helices or parabolas) in three dimensions. As we show, 
sharp upper bounds on the number of such incidences translate back 
to sharp lower bounds on the number of distinct distances. 
Incidence problems in three dimensions between points and curves 
have been studied in several recent works \cite{AKS,EKS,SW}, and 
a major push in this direction has been made last year, with the 
breakthrough result of Guth and Katz~\cite{GK}, who have introduced 
methods from algebraic geometry for studying problems of this kind. 
This has been picked up by the authors~\cite{EKS}, where worst-case 
tight bounds on the number of incidences between points and lines 
in three dimensions (under certain restrictions) have been obtained.

The present paper serves two purposes. First, it studies in detail 
the connection between the distinct distances problem and the 
corresponding 3-dimensional incidence problem. As it turns out, 
there is a lot of interesting
geometric structure behind this reduction, and the paper develops it
in detail. We offer several conjectures on the number of incidences,
and show how, if true, they yield the almost tight worst-case lower 
bound $\Omega(s/\log s)$ on the number of distinct distances.
Unfortunately, so far we have not succeeded in proving these
conjectures. Nevertheless, we have made considerable progress on the
incidence problem itself, which is the second purpose of the study in
this paper. We show how to adapt the algebraic machinery of
\cite{GK,EKS,KSS,Qu} to derive sharp bounds for the incidence problem.
\cite{EKS,GK,KSS,Qu} to derive sharp bounds for the incidence problem.
These bounds are very similar to, and in fact even better than
the bounds obtained in \cite{EKS} for point-line incidences,
where they have been shown to be worst-case tight. However, 
they are not (yet) good enough to yield significant lower bounds 
for distinct distances. We believe that there is additional 
geometric structure in the particular problem studied here, 
which should enable one to further improve the bounds,
but so far this remains elusive.

The paper is organized as follows. We first describe the reduction
from the planar distinct distances problem to the 3-dimensional 
incidence problem mentioned above. In doing so, we note and 
explore several additional geometric connections between the two
problems (as manifested, e.g., in the analysis of {\em special surfaces}
given below). We then present the tools from algebraic geometry that
are needed to tackle the incidence problem; they are variants of the
tools used in \cite{EKS,GK}, adapted to the specific curves that we
need to handle. We then go on to bound the number of incidences.
We first bound the number of rotations in terms of the number of
parabolas, and then bound the number of incidences themselves. 
The latter task is achieved in two steps. We first use a ``purely 
algebraic'' analysis, akin to those in \cite{EKS,GK}, to obtain a 
weaker bound, which we then refine in the second step, using more 
traditional space decomposition techniques. The final bound is 
still not as good as we would like it to be, but it shows that 
the case studied in this paper ``behaves 
better'' than its counterpart involving lines.

\paragraph{Distinct distances and incidences with helices.}
We offer the following novel approach to the problem of distinct
distances.

\paragraph{(H1) Notation.}
Let $S$ be a set of $s$ points in the plane with $x$ distinct
distances. Let $K$ denote the set of all quadruples
$(a,b,a',b')\in S^4$, such that the pairs $(a,b)$ and $(a',b')$
are distinct (although the points themselves need not be) and
$|ab|=|a'b'|>0$.

Let $\delta_1,\ldots,\delta_x$ denote the $x$ distinct distances in
$S$, and let $E_i = \{(a,b)\in S^2 \mid |ab|=\delta_i \}$. We have
$$
|K| = 2\sum_{i=1}^x {|E_i|\choose 2} \ge
\sum_{i=1}^x (|E_i|-1)^2 \ge
\frac{1}{x}\left[ \sum_{i=1}^x (|E_i|-1) \right]^2 =
\frac{\left[s(s-1)-x\right]^2}{x} .
$$

\paragraph{(H2) Rotations.}
We associate each $(a,b,a',b')\in K$ with a (unique) {\em rotation}
(or, rather, a rigid, orientation-preserving
transformation of the plane) $\tau$, 
which maps $a$ to $a'$ and $b$ to $b'$. A rotation $\tau$, in 
complex notation, can be written as the transformation 
$z\mapsto pz+q$, where $p,q\in\comp$ and $|p|=1$. Putting 
$p=e^{i\theta}$, $q=\xi+i\eta$, we can represent $\tau$ by 
the point $(\xi,\eta,\theta) \in \reals^3$. 
In the planar context, $\theta$ is the counterclockwise angle 
of the rotation, and the center of rotation is 
$c=q/(1-e^{i\theta})$, which is defined for $\theta\ne 0$; 
for $\theta=0$, $\tau$ is a pure translation.

The {\em multiplicity} $\mu(\tau)$ of a rotation $\tau$ (with 
respect to $S$) is defined as $|\tau(S)\cap S| =$ the number of 
pairs $(a,b)\in S^2$ such that $\tau(a)=b$. Clearly, one always 
has $\mu(\tau) \le s$, and we will mostly consider only rotations 
satisfying $\mu(\tau)\ge 2$. As a matter of fact, the bulk of the
paper will only consider rotations with multiplicity at least $3$.
Rotations with multiplicity $2$ are harder to analyze.

If $\mu(\tau) = k$ then $S$ contains two congruent and equally
oriented copies $A,B$ of some $k$-element set, such that 
$\tau(A)=B$. Thus, studying multiplicities of rotations is 
closely related to analyzing repeated (congruent and equally 
oriented) patterns in a planar point set; see \cite{BMP} for 
a review of many problems of this kind.

\paragraph{Anti-rotations.}
In this paper we will also consider {\em anti-rotations}, which are
rigid, orientation-reversing transformations of the plane. Any
anti-rotation can be represented as a rotation, followed by a
reflection about some fixed line, e.g., the $x$-axis (so, in 
complex notation, this can be written as 
$z \mapsto \overline{pz+q}$).
Anti-rotations will be useful in certain steps of the analysis.

\paragraph{(H3) Bounding $|K|$.}
If $\mu(\tau) = k$ then $\tau$ contributes $\binom{k}{2}$
quadruples to $K$. Let $N_k$ (resp., $N_{\ge k}$) denote the number
of rotations with multiplicity exactly $k$ (resp., at least $k$),
for $k\ge 2$. Then
$$
|K| = \sum_{k=2}^{s} {k\choose 2} N_k =
  \sum_{k=2}^{s} {k\choose 2} (N_{\ge k} - N_{\ge k+1}) =
N_{\ge 2} + \sum_{k\ge 3} (k-1) N_{\ge k} .
$$

\paragraph{(H4) The main conjecture.} 
\begin{conjecture} \label{conj1}
For any $2\le k\le s$, we have
$$
N_{\ge k} = O\left(s^3/k^2\right).
$$
\end{conjecture}
Suppose that the conjecture were true. Then we would have
$$
\frac{\left[s(s-1)-x\right]^2}{x} \le |K| =
O(s^3) \cdot \left[ 1 + \sum_{k\ge 3} \frac{1}{k} \right] = O(s^3\log s) ,
$$
which would have implied that $x=\Omega(s/\log s)$. This would have 
almost settled the problem of obtaining a tight bound for the minimum
number of distinct distances guaranteed to exist in any set of
$s$ points in the plane, since, as mentioned above, the upper bound 
for this quantity is $O(s/\sqrt{\log s})$ \cite{Er}.

We note that Conjecture~\ref{conj1} is rather deep; even the simple 
instance $k=2$, asserting that there are only $O(s^3)$ rotations 
which map (at least) two points of $S$ to two other points of $S$ 
(at the same distance apart), seems quite difficult. 
In this paper we establish a variety of upper bounds on the number of
rotations and on the sum of their multiplicities. In particular,
these results provide a partial 
positive answer, showing that $N_{\ge 3} = O(s^3)$; that is, the 
number of rotations which map a (degenerate or non-degenerate)
triangle determined by $S$ to another congruent (and equally oriented)
such triangle, is $O(s^3)$. Bounding $N_2$ by $O(s^3)$ is still 
an open problem. See Section~\ref{sec:impr} for a simple proof of the
weaker bound $N_{\ge 2} = O(s^{10/3})$.

\paragraph{Lower bound.}
We next give a construction (suggested by Haim Kaplan)
which shows:
\begin{lemma} \label{lem:lower}
There exist sets $S$ in the plane of arbitrarily large cardinality, 
which determine $\Theta(|S|^3)$ distinct rotations, each mapping
a triple of points of $S$ to another triple of points of $S$.
\end{lemma}

\begin{figure}[htbp]
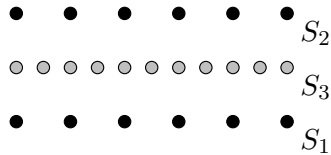

\begin{center}
\input lower.pstex_t
\caption{A lower bound construction of $\Theta(|S|^3)$ rotations with
multiplicity $3$.}
\label{lower}
\end{center}
\end{figure}

\noindent{\bf Proof:}
Consider the set $S=S_1\cup S_2\cup S_3$, where
\begin{eqnarray*}
S_1 & = & \{ (i,0) \mid i=1,\ldots,s \} , \\
S_2 & = & \{ (i,1) \mid i=1,\ldots,s \} , \\
S_3 & = & \{ (i/2,1/2) \mid i=1,\ldots,2s \} .
\end{eqnarray*}
See Figure~\ref{lower}.

For each triple $a,b,c\in\{1,\ldots,s\}$ such that $a+b-c$ also
belongs to $\{1,\ldots,s\}$, construct the rotation $\tau_{a,b,c}$
which maps $(a,0)$ to $(b,0)$ and $(c,1)$ to $(a+b-c,1)$. Since
the distance between the two source points is equal to the distance
between their images, $\tau_{a,b,c}$ is well (and uniquely) defined.
Moreover, $\tau_{a,b,c}$ maps the midpoint $((a+c)/2,1/2)$ to the
midpoint $((a+2b-c)/2,1/2)$.

We claim that the rotations $\tau_{a,b,c}$ are all distinct. Indeed,
suppose that two such rotations, $\tau_{a,b,c}$ and $\tau_{a',b',c'}$,
for distinct triples $(a,b,c)$, $(a',b',c')$, coincide; call the
common rotation $\tau$. We can represent $\tau$ as the rigid 
transformation which first translates the plane horizontally by 
distance $b-a$, so that $(a,0)$ is mapped to $(b,0)$, and then 
rotates it around $(b,0)$ by an appropriate angle $0<\theta<\pi$, 
so that $(c+b-a,1)$ is mapped to $(a+b-c,1)$. Suppose first that 
$a\ne a'$. Since $\tau = \tau_{a,b,c} = \tau_{a',b',c'}$, it maps 
$(a',0)$ to $(a'+b-a,0)$ and then rotates this point by angle 
$\theta$ around $(b,0)$, mapping it to a point outside the 
$x$-axis, contradicting the fact that $\tau_{a',b',c'}$ maps 
$(a',0)$ to $(b',0)$. If $a'=a$ then we also must have $b'=b$, 
so $c'\ne c$. But then it is impossible to turn, around $(b,0)$, the
shifted point $(c+b-a,1)$ to $(a+b-c,1)$ and the
shifted point $(c'+b-a,1)$ to $(a+b-c',1)$, by the same angle, a
contradiction which shows that the two rotations are distinct.

Since there are $\Theta(s^3)$ triples $(a,b,c)$ with the above
properties, the claim follows.
$\Box$

\noindent{\bf Remarks.}
{\bf (1)} A ``weakness'' of this construction is that all the rotations
$\tau_{a,b,c}$ map a {\em collinear} triple of points of $S$ to
another collinear triple. (In the terminology to follow, these will be
called {\em flat} rotations.) We do not know whether the number of
rotations which map a {\em non-collinear} triple of points of $S$ 
to another non-collinear triple can be $\Omega(|S|^3)$.
We tend to conjecture that this is indeed the case.

\noindent
{\bf (2)} We do not know whether Conjecture~\ref{conj1} 
is worst-case tight (if true). 
That is, we do not know whether there exist sets $S$, with 
$s=|S|$ arbitrarily large, so that there are $\Omega(s^3/k^2)$
distinct rotations, each mapping at least $k$ points of $S$ to 
$k$ other points of $S$.

\paragraph{(H5) Helices.}
To estimate $N_{\ge k}$, we reduce the problem of analyzing
rotations and their interaction with $S$ to an incidence problem
in three dimensions, as follows.

With each pair $(a,b)\in S^2$, we associate the curve $h_{a,b}$,
in a 3-dimensional space parametrized by $(\xi,\eta,\theta)$, 
which is the locus of all rotations which map $a$ to $b$. 
That is, the equation of $h_{a,b}$ is given by
$$
h_{a,b} = \{ (\xi,\eta,\theta) \mid b = ae^{i\theta} + (\xi,\eta) \}.
$$
Putting $a=(a_1,a_2)$, $b=(b_1,b_2)$, this becomes
\begin{eqnarray} \label{eq:helix}
\xi & = & b_1 - (a_1\cos\theta - a_2\sin\theta) , \\
\eta & = & b_2 - (a_1\sin\theta + a_2\cos\theta) \nonumber .
\end{eqnarray}
This is a {\em helix} in $\reals^3$, having four degrees of freeedom,
which we parametrize by $(a_1,a_2,b_1,b_2)$. It extends from the plane
$\theta=0$ to the plane $\theta=2\pi$; its two endpoints lie
vertically above each other, and it completes exactly one 
revolution between them.

\paragraph{(H6) Helices, rotations, and incidences.}
Let $P$ be a set of rotations, represented by points in $\reals^3$,
as above, and let $H$ denote the set of all $s^2$ helices
$h_{a,b}$, for $(a,b)\in S^2$ (note that $a=b$ is permitted). 
Let $I(P,H)$ denote the number of incidences between $P$ and $H$. 
Then we have
$$
I(P,H) = \sum_{\tau\in P} \mu(\tau) .
$$
Rotations $\tau$ with $\mu(\tau)=1$ are not interesting, because 
each of them only contributes $1$ to the count $I(P,H)$, and we will
mostly ignore them. For the same reason, rotations with 
$\mu(\tau)=2$ are also not interesting for estimating $I(P,H)$, 
but they need to be included in the analysis of $N_{\ge 2}$. 
Unfortunately, as already noted, we do not yet have a good 
upper bound (i.e., cubic in $s$) on the number of such rotations.

\paragraph{(H7) Incidences and the second conjecture.} 

\begin{conjecture} \label{conj2}
For any $P$ and $H$ as above, we have
$$
I(P,H) = O\left(|P|^{1/2}|H|^{3/4}+|P|+|H|\right) .
$$
\end{conjecture} 

Suppose that Conjecture~\ref{conj2} were true. 
Let $P_{\ge k}$ denote the set of all rotations with 
multiplicity at least $k$ (with respect to $S$). We then have
$$
kN_{\ge k} = k|P_{\ge k}| \le I(P_{\ge k},H) =
O\left(N_{\ge k}^{1/2}|H|^{3/4} + N_{\ge k} + |H|\right) ,
$$
from which we obtain
$$
N_{\ge k} = O\left( \frac{s^3}{k^2} + \frac{s^2}{k} \right) =
O\left( \frac{s^3}{k^2} \right) ,
$$
thus establishing Conjecture~\ref{conj1}, and therefore also the lower
bound for $x$ (the number of distinct distances) 
derived above from this conjecture.

\noindent{\bf Remark.}
Conjecture~\ref{conj2} can also be formulated for an {\em arbitrary}
subset $H$ of all possible helices.

Note that two helices $h_{a,b}$ and $h_{c,d}$ intersect in at most one
point---this is the unique rotation which maps $a$ to $b$ and $c$ to
$d$ (if it exists at all, namely if $|ac|=|bd|$). Hence, combining
this fact with a standard cutting-based decomposition technique, 
similar to what has been noted in \cite{SW}, say, yields the 
weaker bound
\begin{equation} \label{weak23}
I(P,H) = O\left(|P|^{2/3}|H|^{2/3}+|P|+|H|\right) ,
\end{equation} 
which, alas, only yields the much weaker bound
$N_{\ge k} = O\left( s^4/k^3 \right)$,
which is completely useless for deriving any lower bound on $x$.
(We will use this bound, though, in Section~\ref{sec:conc}.)

\paragraph{(H8) From helices to parabolas.}
The helices $h_{a,b}$ are non-algebraic curves, because of the 
use of the angle $\theta$ as a parameter. This can be easily 
remedied, in the following standard manner.
Assume that $\theta$ ranges from $-\pi$ to $\pi$, and
substitute, in the equations (\ref{eq:helix}), 
$Z=\tan(\theta/2)$, 
$X = \xi(1+Z^2)$, and $Y = \eta(1+Z^2)$, to obtain
\begin{eqnarray} \label{parabola}
X & = & (a_1+b_1)Z^2 + 2a_2Z + (b_1-a_1) \\
Y & = & (a_2+b_2)Z^2 - 2a_1Z + (b_2-a_2) , \nonumber 
\end{eqnarray}
which are the equations of a {\em planar parabola} in the
$(X,Y,Z)$-space.
(The parabola degenerates to a line if $b=-a$, a situation that we
will rule out by choosing an appropriate generic coordinate frame
in the original $xy$-plane.)
We denote the parabola corresponding to the helix $h_{a,b}$ as
$h^*_{a,b}$, and refer to it as an {\em $h$-parabola}.

\paragraph{(H9) Joint and flat rotations.}
A rotation $\tau\in P$ is called a {\em joint} of $H$ if $\tau$ is
incident to at least three helices of $H$ whose tangent lines at 
$\tau$ are non-coplanar. Otherwise, still assuming that $\tau$ is
incident to at least three helices of $H$, $\tau$ is called {\em flat}.

Let $\tau=(\xi,\eta,\theta)\in P$ be a rotation, incident 
to three distinct helices $h_{a,b}$, $h_{c,d}$, $h_{e,f}$.
From their equations, as given in (\ref{eq:helix}),
the directions of the tangents to these helices at $\tau$ are
\begin{align*}
&
(a_1\sin\theta + a_2\cos\theta,\, -a_1\cos\theta + a_2\sin\theta,\, 1)
\\
&
(c_1\sin\theta + c_2\cos\theta,\, -c_1\cos\theta + c_2\sin\theta,\, 1)
\\
&
(e_1\sin\theta + e_2\cos\theta,\, -e_1\cos\theta + e_2\sin\theta,\, 1) .
\end{align*}
Put $p=\cos\theta$ and $q=\sin\theta$. Then the three tangents are
coplanar if and only if
$$
\left|
\begin{array}{ccc}
a_1q+a_2p & -a_1p+a_2q & 1 \\
c_1q+c_2p & -c_1p+c_2q & 1 \\
e_1q+e_2p & -e_1p+e_2q & 1 
\end{array}
\right| = 0 .
$$
Simplifying the determinant, and recalling that $p^2+q^2=1$,
the condition is equivalent to
$$
\left|
\begin{array}{ccc}
a_1 & a_2 & 1 \\
c_1 & c_2 & 1 \\
e_1 & e_2 & 1 
\end{array}
\right| = 0 .
$$
In other words, the three helices $h_{a,b}$, $h_{c,d}$, $h_{e,f}$ form
a joint at $\tau$ if and only if the three points $a,c,e$ (and thus
also $b,d,f$) are non-collinear.  That is, we have shown:

\begin{claim} \label{jointri}
A rotation $\tau$ is a joint of $H$ if and only if $\tau$ maps a
non-degenerate triangle determined by $S$ to another (congruent
and equally oriented) non-degenerate triangle determined by $S$.
A rotation $\tau$ is a flat rotation if and only if $\tau$ maps at
least three collinear points of $S$ to another collinear triple of
points of $S$, but
does not map any point of $S$ outside the line containing the triple
to another point of $S$.
\end{claim}

\noindent{\bf Remarks:}
{\bf (1)} Note that if $\tau$ is a flat rotation, it maps the entire line
containing the three source points to the line containing their
images. Specifically (see also below), we can respectively
parametrize points on these lines as $a_0+tu$, $b_0+tv$, 
for $t\in\reals$, such that 
$\tau$ maps $a_0+tu$ to $b_0+tv$ for every $t$.

\noindent
{\bf (2)} For flat rotations, we also need to ensure, for technical
reasons, that the three (or more) 
helices incident to a flat rotation $\tau$ are such that their 
tangents at $\tau$ are all distinct. This fortunately is always 
the case. Indeed, The preceding analysis is easily seen to imply 
that if $h_{a,b}$ and $h_{c,d}$ meet at $\tau$ then their tangents at
$\tau$ coincide if and only if $a=c$. But then $h_{a,b}$ and $h_{a,d}$
cannot have a common point (rotation) unless $b=d$ too, i.e.,
they are the same helix; otherwise the common rotation would have to
map $a$ to the two distinct points $b$ and $d$, an impossibility.

\paragraph{(H10) Special surfaces.}
In preparation for the forthcoming algebraic analysis, we need the
following property of our helices.

Let $\tau$ be a flat rotation, with multiplicity $k\ge 3$,
and let $\ell$ and $\ell'$ be the corresponding lines in the plane,
such that there exist $k$ points $a_1,\ldots,a_k\in S\cap \ell$ 
and $k$ points $b_1,\ldots,b_k\in S\cap \ell'$, such that $\tau$ 
maps $a_i$ to $b_i$ for each $i$ (and in particular maps $\ell$ 
to $\ell'$). By definition, $\tau$ is incident to the $k$
helices $h_{a_i,b_i}$, for $i=1,\ldots,k$. 

Let $u$ and $v$ denote unit vectors in the direction of $\ell$ and
$\ell'$, respectively. Clearly, there exist two reference points
$a\in\ell$ and $b\in\ell'$, such that for each $i$ there is a 
real number $t_i$ such that $a_i=a+t_iu$ and $b_i=b+t_iv$. 
As a matter of fact, for each real $t$, $\tau$ maps $a+tu$ to
$b+tv$, so it is incident to $h_{a+tu,b+tv}$.
Note that $a$ and $b$ are not uniquely defined: we can take $a$ to 
be any point on $\ell$, and shift $b$ accordingly along $\ell'$.

Let $H(a,b;u,v)$ denote the set of these helices.
Since a pair of helices can meet in at most one point, all the helices
in $H(a,b;u,v)$ pass through $\tau$ but are otherwise pairwise disjoint.
Using the re-parametrization $(\xi,\eta,\theta)\mapsto (X,Y,Z)$, 
we denote by $\Sigma=\Sigma(a,b;u,v)$ the surface which is the 
union of all the $h$-parabolas that are the images of the helices 
in $H(a,b;u,v)$. 
We refer to such a surface $\Sigma$ as a {\em special surface}.

An important comment is that most of the ongoing analysis also applies 
when only two helices are incident to $\tau$; they suffice to determine 
the four parameters $a,b,u,v$ that define the surface $\Sigma$.

We also remark that, although we started the definition of
$\Sigma(a,b;u,v)$ with a flat rotation $\tau$, the definition only 
depends on the parameters $a,b,u$, and $v$ (and even there we have,
as just noted, one degree of freedom in choosing $a$ and $b$). 
If $\tau$ is not flat
it may determine many special surfaces, one for each line that contains
two or more points of $S$ which $\tau$ maps to other (also collinear)
points of $S$. Also, as we will shortly see, the same surface can 
be obtained from a different set (in fact, many such sets) of 
parameters $a',b',u'$, and $v'$ (or, alternatively, from different 
flat rotations $\tau'$). 
An ``intrinsic'' definition of special surfaces will be given shortly.

The surface $\Sigma$ is a cubic algebraic surface, whose
equation can be worked out as follows.
The equation of the parabola $h^*_{a+tu,b+tv}$ corresponding
to $h_{a+tu,b+tv}$ is
\begin{eqnarray*}
X & = & (a_1+b_1+t(u_1+v_1))Z^2 + 2(a_2+tu_2)Z + (b_1-a_1+t(v_1-u_1)) \\
Y & = & (a_2+b_2+t(u_2+v_2))Z^2 - 2(a_1+tu_1)Z + (b_2-a_2+t(v_2-u_2)) .
\end{eqnarray*}
We can view this as a parametrization of $\Sigma$ using $t$ and $Z$ as
parameters. We can simplify these equations as
\begin{eqnarray} \label{parsig}
X & = & tQ_1(Z) + Q_3(Z) \\
Y & = & tQ_2(Z) + Q_4(Z) , \nonumber
\end{eqnarray}
where $Q_1,\ldots,Q_4$ are quadratic polynomials in $Z$. Eliminating
$t$ from these equations gives us the first version of the
equation of $\Sigma$, which is
\begin{equation} \label{sigma1}
Q_2(Z)X - Q_1(Z)Y + (Q_1(Z)Q_4(Z)-Q_2(Z)Q_3(Z)) = 0 .
\end{equation}
This is a quartic equation, although it is only linear in $X$ and $Y$.

Note also that the cross-section of $\Sigma$ by any plane 
$Z={\rm const}$ is a line, so $\Sigma$ is a ruled surface.

We next reduce (\ref{sigma1}) to a cubic equation, as follows.
Let $(X_0,Y_0,Z_0)$ denote the coordinates of $\tau$ in the
$XYZ$-frame. We note that $Q_1(Z_0)=Q_2(Z_0)=0$. 
This can be worked out 
explicitly, or concluded by noting that $(X_0,Y_0,Z_0)$ is a 
common point of all our parabolas, so $(X_0,Y_0,Z_0)$ cannot
determine $t$, meaning that the coefficients $Q_1(Z_0)$ and 
$Q_2(Z_0)$ in (\ref{parsig}) must both be zero.

Hence, each of the three polynomials $Q_2$, $Q_1$, and
$Q_1Q_4-Q_2Q_3$, appearing in the left-hand side of 
(\ref{sigma1}), vanishes at $Z_0$, and is therefore 
divisible by $Z-Z_0$. Factoring $Z-Z_0$ out,
we get a reduced equation for $\Sigma$, of the form
\begin{equation} \label{sigma}
E_2(Z)X - E_1(Z)Y + (E_1(Z)Q_4(Z)-E_2(Z)Q_3(Z)) = 0 ,
\end{equation} 
where $E_1$ and $E_2$ are linear in $Z$. 
Recalling that
\begin{eqnarray*}
Q_1(Z) & = & (u_1+v_1)Z^2 + 2u_2Z + (v_1-u_1) \\
Q_2(Z) & = & (u_2+v_2)Z^2 - 2u_1Z + (v_2-u_2) \\
Q_3(Z) & = & (a_1+b_1)Z^2 + 2a_2Z + (b_1-a_1) \\
Q_4(Z) & = & (a_2+b_2)Z^2 - 2a_1Z + (b_2-a_2) ,
\end{eqnarray*}
an explicit calculation yields:
\begin{eqnarray*}
E_1(Z) & = & (u_1+v_1)(Z+Z_0) + 2u_2 \\
E_2(Z) & = & (u_2+v_2)(Z+Z_0) - 2u_1 .
\end{eqnarray*}
An additional explicit calculation shows that 
\begin{equation} \label{e12z0}
E_1(Z_0) = 2v_2 \quad\quad\mbox{and}\quad\quad E_2(Z_0) = -2v_1 .
\end{equation}
(To see, say, the first equality, we need to show that
$(u_1+v_1)Z_0 = v_2-u_2$. Writing $u=(\cos\alpha,\sin\alpha)$,
$v=(\cos(\alpha+\theta),\sin(\alpha+\theta))$, where $\theta$ 
is the angle of rotation, and recalling that
$Z_0=\tan\frac{\theta}{2}$, the claim follows by straightforward
trigonometric manipulations.)

This allows us to rewrite
\begin{eqnarray} \label{e1e2}
E_1(Z) & = & (u_1+v_1)Z + (u_2+v_2) \\
E_2(Z) & = & (u_2+v_2)Z - (u_1+v_1) \nonumber .
\end{eqnarray}
Hence, the ``free'' term in (\ref{sigma}) is the cubic polynomial
$$
E_1(Z)Q_4(Z)-E_2(Z)Q_3(Z) = 
$$
$$
\biggl( (u_1+v_1)Z + (u_2+v_2) \biggr)
\biggl( (a_2+b_2)Z^2 - 2a_1Z + (b_2-a_2) \biggr) -
$$
$$
\biggl( (u_2+v_2)Z - (u_1+v_1) \biggr)
\biggl( (a_1+b_1)Z^2 + 2a_2Z + (b_1-a_1) \biggr) .
$$
We refer to the cubic polynomial in the left-hand side of
(\ref{sigma}) as a {\em special polynomial}. Thus a special surface is
the zero set of a special polynomial.


\paragraph{(H11) The geometry of special surfaces.}
Special surfaces pose a technical challenge to the analysis.
Specifically, each special surface $\Sigma$ captures a certain
underlying pattern in the ground set $S$, which may result in many
incidences between rotations and $h$-parabolas, all contained in
$\Sigma$. The next step of the analysis studies this pattern in
detail. 

\begin{figure}[htbp]
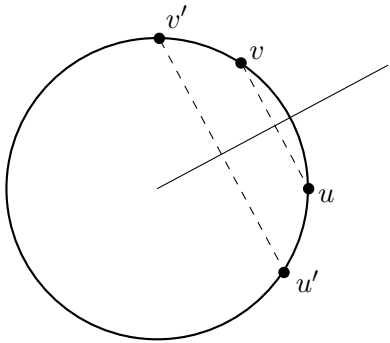

\begin{center}
\input uvuv1.pstex_t
\caption{The configuration of $u,v,u',v'$.}
\label{uvuv1}
\end{center}
\end{figure}

\begin{figure}[htbp]
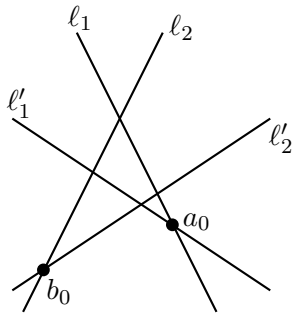

\begin{center}
\input uvuv2.pstex_t
\caption{The structure of $\tau$ and $\tau'$ on a common special
surface $\Sigma$.}
\label{uvuv2}
\end{center}
\end{figure}

Consider first a simple instance of this situation, in which 
two special surfaces $\Sigma$, $\Sigma'$,
generated by two distinct flat rotations $\tau$, $\tau'$, coincide.
More precisely, there exist four parameters $a,b,u,v$ such that 
$\tau$ maps the line $\ell_1 = a+tu$ to the line $\ell_2 = b+tv$
(so that points with the same parameter $t$ are mapped to one 
another), and four other parameters $a',b',u',v'$ such that 
$\tau'$ maps (in a similar manner) the line $\ell'_1 = a'+tu'$ 
to the line $\ell'_2 = b'+tv'$, and
$\Sigma(a,b;u,v) = \Sigma(a',b';u',v')$. Denote this common 
surface by $\Sigma$.
Since the surfaces coincide, the coefficients $E_1(Z)$, $E_2(Z)$ 
for $(a,b,u,v)$ must be proportional to the coefficients $E'_1(Z)$, 
$E'_2(Z)$ for $(a',b',u',v')$.  That is, we must have 
$u'_1+v'_1 = \gamma(u_1+v_1)$ and $u'_2+v'_2 = \gamma(u_2+v_2)$, 
for some real $\gamma$. In other words, $u'+v' = \gamma(u+v)$. 
Since $u,v,u',v'$ are unit vectors, 
the angle bisector between $u$ and $v$ must coincide with that 
between $u'$ and $v'$, as depicted in Figure~\ref{uvuv1}. Moreover, 
as is easily checked, 
if we let $a_0$ be the intersection point of $\ell_1$ and $\ell'_1$, 
and let $b_0$ be the intersection point of $\ell_2$ and $\ell'_2$, 
then both $\tau$ and $\tau'$ map $a_0$ to $b_0$, and $h^*_{a_0,b_0}$
is contained in $\Sigma$.  (See Figure~\ref{uvuv2}.)
Indeed, $\tau'$ lies on some parabola
$h^*_{p,q}$ through $\tau$ which is contained in $\Sigma$, and
$\tau$ lies on some parabola $h^*_{p',q'}$ through $\tau'$ which 
is also contained in $\Sigma$. Since a pair of distinct 
$h$-parabolas meet in at most one point, the two parabolas must
coincide, so $p=p'$ and $q=q'$. However, by construction, $p$ lies on
$\ell_1$ and $p'$ lies on $\ell'_1$, so this common point must be
$a_0$, and, similarly, $q=q'=b_0$, as claimed.

Since the preceding analysis applies to any pair of distinct 
rotations on a common special surface $\Sigma$, it follows that
we can associate with $\Sigma$ a common direction $w$ and a common
shift $\delta$, so that for each $\tau\in\Sigma$ there exist two lines
$\ell$, $\ell'$, where $\tau$ maps $\ell$ to $\ell'$, so that the
angle bisector between these lines is in direction $w$, and $\tau$ is
the unique rigid motion, obtained by rotating $\ell$ to $\ell'$ 
around their intersection point $\ell\cap\ell'$, and then
shifting $\ell'$ along itself by a distance whose projection in
direction $w$ is $\delta$. The fact that the shifts of any pair of
rotations on $\Sigma$ have the same $w$-component follows from 
the fact that they both map the intersection point $a_0$ of their
source lines to the intersection point $b_0$ of their target lines;
consult Figure~\ref{uvuv2}.

Let $\Sigma$ be a special surface, generated by $H(a,b;u,v)$; that is,
$\Sigma$ is the union of all parabolas of the form $h^*_{a+tu,b+tv}$,
for $t\in\reals$. Let $\tau_0$ be the common rotation to all these
parabolas, so it maps the line $\ell_0 = \{a+tu \mid t\in\reals\}$
to the line $\ell'_0 = \{b+tv \mid t\in\reals\}$, so that every point
$a+tu$ is mapped to $b+tv$.

\begin{figure}[htbp]
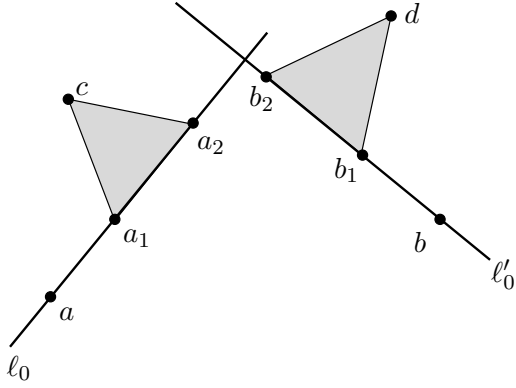

\begin{center}
\input uvuv3.pstex_t
\caption{The geometric configuration corresponding to a parabola
$h^*_{c,d}$ contained in $\Sigma$.}
\label{uvuv3}
\end{center}
\end{figure}

Let $h^*_{c,d}$ be a parabola contained in $\Sigma$ but not passing
through $\tau_0$. Take any pair of distinct rotations 
$\tau_1,\tau_2$ on $h^*_{c,d}$. Then there exist two respective real
numbers $t_1,t_2$, such that $\tau_i\in h^*_{a+t_iu,b+t_iv}$, for
$i=1,2$. Thus $\tau_i$ is the unique rotation which maps $c$ to $d$
and $a_i=a+t_iu$ to $b_i=b+t_iv$. In particular, we have
$|a+t_iu-c| = |b+t_iv-d|$. This in turn implies that the triangles 
$a_1a_2c$ and $b_1b_2d$ are congruent; see Figure~\ref{uvuv3}.

Given $c$, this determines $d$, up to a reflection about $\ell'_0$. We
claim that $d$ has to be on the ``other side'' of $\ell'_0$, namely,
be such that the triangles $a_1a_2c$ and $b_1b_2d$ are oppositely oriented.
Indeed, if they were equally oriented, then $\tau_0$ would have mapped
$c$ to $d$, and then $h^*_{c,d}$ would have passed through $\tau_0$,
contrary to assumption.

Now form the two sets
\begin{eqnarray} \label{aandb}
A & = & \{ p \mid \mbox{there exists $q\in S$ such that}\;
h^*_{p,q} \subset \Sigma \} \\
B & = & \{ q \mid \mbox{there exists $p\in S$ such that}\;
h^*_{p,q} \subset \Sigma \} . \nonumber
\end{eqnarray}
The preceding discussion implies that $A$ and $B$ are congruent and 
oppositely oriented. 

To recap, each rotation $\tau\in\Sigma$, incident to $k\ge 2$
parabolas contained in $\Sigma$, corresponds to a pair of lines 
$\ell,\ell'$ with the above properties, so that $\tau$ maps $k$ 
points of $S\cap\ell$ (rather, of $A\cap\ell$) to $k$ points of
$S\cap\ell'$ (that is, of $B\cap\ell'$).
If $\tau$ is flat, its entire multiplicity comes 
from points of $S$ on $\ell$ (these are the points
of $A\cap\ell$) which are mapped by $\tau$ to points of $S$ on 
$\ell'$ (these are points of $B\cap\ell'$), and all the corresponding
parabolas are contained in $\Sigma$. If $\tau$ is a joint then, 
for any other point $p\in S$ outside $\ell$ which is mapped by 
$\tau$ to a point $q\in S$ outside $\ell'$, the parabola 
$h^*_{p,q}$ is not contained in $\Sigma$, and crosses it 
transversally at the unique rotation $\tau$.

Note also that any pair of parabolas $h^*_{c_1,d_1}$ and 
$h^*_{c_2,d_2}$ which are contained in $\Sigma$ intersect, 
necessarily at the unique rotation which maps $c_1$ to $d_1$ and 
$c_2$ to $d_2$. This holds because $|c_1c_2|=|d_1d_2|$, as follows
from the preceding discussion.

\paragraph{Special surfaces are anti-rotations.}
Let $\Sigma$ be a special surface, and let $A,B$ be the subsets of
$S$ associated with $\Sigma$, as in (\ref{aandb}).
Then there exists a single {\em anti-rotation} which maps $A$ to $B$.
Conversely, any anti-rotation can be associated with a unique 
special surface in this manner. However, the number
of incidences within a special surface may be larger than the
incidence count of the anti-rotation with the appropriate variants of
the $h$-parabolas: the former counts incidences between the points of
$A$ (or of $B$) and the lines that they determine, while the latter 
only counts the size of $A$ (or of $B$).

\paragraph{An alternative analysis.}
Recall the equation (\ref{sigma}) of $\Sigma$
$$
E_2(Z)X - E_1(Z)Y + (E_1(Z)Q_4(Z)-E_2(Z)Q_3(Z)) = 0 ,
$$
where, writing $\lambda = u_1+v_1$ and $\mu = u_2+v_2$,
\begin{eqnarray*} 
E_1(Z) & = & \lambda Z + \mu \\
E_2(Z) & = & \mu Z - \lambda .
\end{eqnarray*}
Now let $h^*_{a,b}$ be a parabola contained in $\Sigma$. Substituting 
the equations (\ref{parabola}) of $h^*_{a,b}$ into the above equation, we
get
$$
(\mu Z - \lambda) \biggl[ 
(a_1+b_1)Z^2 + 2a_2Z + (b_1-a_1) 
\biggr]
- (\lambda Z + \mu) \biggl[
(a_2+b_2)Z^2 - 2a_1Z + (b_2-a_2) 
\biggr] + K(Z) \equiv 0 ,
$$
where $K(Z) = E_1(Z)Q_4(Z)-E_2(Z)Q_3(Z)$ is the ``free'' cubic
term in the equation of $\Sigma$.
A straightforward algebraic simplification of this equation yields
$$
(Z^2+1) \biggl[
(\mu Z + \lambda) a_1 -
(\lambda Z - \mu) a_2 +
(\mu Z - \lambda) b_1 -
(\lambda Z + \mu) b_2 
\biggr] + K(Z) \equiv 0 .
$$
In particular (an interesting observation in itself, albeit obvious 
from the definition of $X,Y,Z$), $K(Z)$ must be divisible by 
$Z^2+1$, with the remainder being a linear function of $Z$. 
Eliminating this factor, we get 
\begin{eqnarray*}
\mu (a_1+b_1) - \lambda (a_2+b_2) & = & c_1 \\
\lambda (a_1-b_1) + \mu (a_2-b_2) & = & c_2 ,
\end{eqnarray*}
for appropriate reals numbers $c_1$, $c_2$.

Now, writing $u = (\cos\alpha,\sin\alpha)$ and
$v = (\cos(\alpha+\theta),\sin(\alpha+\theta))$, where $\theta$ is the
angle of rotation, and observing that
$$
u+v = (u_1+v_1,u_2+v_2) = (\lambda,\mu) = 
\cos\tfrac{\theta}{2} \left(
\cos \left(\alpha+\tfrac{\theta}{2}\right),\;
\sin \left(\alpha+\tfrac{\theta}{2}\right)
\right) ,
$$
the containment of $h^*_{a,b}$ in $\Sigma$ is equivalent to the two
conditions
\begin{eqnarray*}
(a+b)\cdot (u+v)^T & = & c'_1 \\
(a-b)\cdot (u+v) & = & c'_2 ,
\end{eqnarray*}
for appropriate parameters $c'_1,c'_2$.
The geometric interpretation of the first condition is that the midpoint of
$ab$ has to lie on a fixed line $\ell_0$ (whose direction,
$\alpha+\frac{\theta}{2}$, is parallel to the angle bisector
between the lines $\ell_1,\ell_2$ (see Figure~\ref{uvuv2}).
The second condition means that $b-a$ has a fixed component in the
direction of $\ell_0$. In other words, $h^*_{a,b}$ is contained in 
$\Sigma$ if and only if $b=\varphi(a)$, where $\varphi$ is the
anti-rotation obtained as a reflection about $\ell_0$ followed 
by a shift parallel to $\ell_0$. This constitutes an alternative 
derivation of the characterization of $\Sigma$ given above.

\paragraph{(H12) Special surfaces and parabolas.}
Finally, we study intersection patterns involving special surfaces.
Let $\Sigma$ be a special surface as above, and let
$\Xi$ be another $(X,Y)$-linear surface of the form
$A(Z)X+B(Z)Y+C(Z)=0$. Then either $\Xi$ coincides with $\Sigma$,
or there is at most one parabola contained in both of them.
Indeed, the intersection of $\Xi$ and $\Sigma$ is the curve 
satisfying
\begin{eqnarray*}
A(Z)X+B(Z)Y+C(Z) & = & 0 \\
E_2(Z)X - E_1(Z)Y + (E_1(Z)Q_4(Z)-E_2(Z)Q_3(Z)) & = & 0 .
\end{eqnarray*}
This is a linear system in $X$ and $Y$. Suppose first that its 
determinant, $A(Z)E_1(Z) + B(Z)E_2(Z)$, does not vanish identically. 
Then, with the exception of finitely many values of $Z$, we get a 
unique solution of the form $X=F(Z)$, $Y=G(Z)$, which can describe 
at most one parabola. If the determinant vanishes identically, then 
the equation of $\Xi$ can be written as $E_2(Z)X-E_1(Z)Y+D(Z)=0$, 
for an appropriate rational algebraic function $D(Z)$. 
If $\Xi$ and $\Sigma$ do intersect in a parabola, then we must have
$D(Z)\equiv E_1(Z)Q_4(Z)-E_2(Z)Q_3(Z)$, so $\Xi$ and $\Sigma$ coincide.
$\Box$

As a corollary, we have:
\begin{lemma} \label{only1}
Let $\Xi$ be an $(X,Y)$-linear surface of the above 
form, and let $\tau$ be a flat rotation contained in $\Xi$. Then either
$\Xi$ contains at least two of the parabolas incident to $\tau$, and 
then it must coincide with the corresponding special surface $\Sigma$, or 
$\Xi$ contains at most one of these parabolas, so at least two other 
parabolas cross $\Xi$ at $\tau$.
\end{lemma}

\begin{lemma} \label{sspec}
A special surface can contain at most $s$ $h$-parabolas.
\end{lemma}

\noindent{\bf Proof:}
Let $\Xi$ be the given special surface. We claim that for each 
$a\in S$ there can be at most one point $b\in S$ such that
$h^*_{a,b}\subset\Xi$. Indeed, suppose that there exist two such points
$b_1,b_2\in S$. Since any pair of $h$-parabolas on $\Xi$ intersect,
$h^*_{a,b_1}$ and
$h^*_{a,b_2}$ meet at a rotation $\tau$, which maps $a$ to both $b_1$
and $b_2$, an impossibility which completes the proof.
$\Box$

\begin{lemma} \label{surfpar}
The number of containments between $n$ $h$-parabolas and $E$ 
special surfaces is 
$$
O(E^{2/3}n^{2/3}+E+n) .
$$
\end{lemma}

\noindent{\bf Proof:}
As argued above, a special surface $\Sigma$ is characterized by
an anti-rotation $\varphi_\Sigma$ in the plane, specified
by a line $\ell$ and a shift $\delta$, 
such that $\varphi_\Sigma(a)$ is the point obtained by reflecting 
$a$ about $\ell$ and then by shifting the reflected point parallel 
to $\ell$ by distance $\delta$. Thus $\Sigma$ has three degrees of 
freedom, and can be parametrized by $(\alpha,\beta,\delta)$, where 
$y=\alpha x+\beta$ is the equation of $\ell$ and $\delta$ is the shift.
We write $\Sigma(\alpha,\beta,\gamma)$ to denote the special surface
parametrized by $(\alpha,\beta,\gamma)$.

By construction, a parabola $h^*_{a,b}$ is contained in $\Sigma$ 
if and only if $\varphi_\Sigma(a) = b$.

We use the following parametric setup. We represent each special 
surface $\Sigma$ by the corresponding triple 
$(\alpha,\beta,\delta)$, and regard it as a 
point in parametric 3-space. Each parabola $h^*_{a,b}$ is mapped
to the locus $\tilde{h}_{a,b}$ of all (points representing) 
special surfaces containing $h^*_{a,b}$. This is a curve in the
$(\alpha,\beta,\delta)$-space, given by the pair of scalar 
equations $\varphi_{\Sigma(\alpha,\beta,\delta)}(a) = b$.
This is a low-degree algebraic curve, whose concrete equations 
can be worked out explicitly, but we skip over this step.

We thus have a system of $E$ points and $n$ such curves in 3-space,
and we wish to bound the number of incidences between them. We have
the additional property, noted in Lemma~\ref{only1}, that two
curves meet in at most one point. By projecting these points and
curves onto some generic 2-plane, one can easily show that that 
the number of incidences, and thus the number of original
containments, is at most $O(E^{2/3}n^{2/3}+E+n)$, as claimed.
$\Box$

\noindent{\bf Remark.}
If we represent each special surface by its corresponding
anti-rotation, Lemma~\ref{surfpar} simply bounds the number 
of incidences between $E$ anti-rotations and $n$ (appropriately
transformed copies of) $h$-parabolas, and the bound noted in
(\ref{weak23}) holds here as well.

\section{Tools from algebraic geometry}
\label{sec:tools}

We begin by reviewing and extending the basic tools from algebraic
geometry which have been used in \cite{GK} and in \cite{EKS}.
However, we develop them here in the context of incidences between
points and our $h$-parabolas, rather than the context of
points and lines considered in the previous papers.

So let $C$ be a set of $n\le s^2$ $h$-parabolas in $\reals^3$.
For each $h^*\in C$, we
denote the plane containing $h^*$ by $\pi_{h^*}$ and its equation
as $L_{h^*}=0$, where $L_{h^*}$ is a linear polynomial. We represent
$h^*$ as the intersection curve of $L_{h^*}=0$ and $F_{h^*}=0$, 
where $F_{h^*}$ is one of the quadratic equations in (\ref{parabola})
defining $h^*$, say the first one.

Note that all the parabolas of $C$ cross every plane of the form 
$Z={\rm const}$, each at a single point. 

Recalling the definitions in (H9), and similar to
the case of lines, we say that a point\footnote{%
  Recall that points in 3-space represent rotations in the plane.
  Later on we will mostly refer to them as rotations, but in the 
  more abstract algebraic treatment in this section
  we prefer to call them points.}
$a$ is a {\em joint} of $C$ if it is incident to three parabolas of
$C$ whose tangents at $a$ are non-coplanar. 
Let $J=J_C$ denote the set of joints of $C$.
We will also consider points $a$ that are incident to three or more 
parabolas of $C$, so that the tangents to all these parabolas are 
coplanar, and refer to such points as {\em flat} points of $C$.
We recall (see (H9)) that any pair of distinct $h$-parabolas 
which meet at a point have there distinct tangents.

First, we note that a trivariate polynomial $p$ of degree $d$ which 
vanishes at $2d+1$ points that lie on a common parabola $h^*\in C$ 
must vanish identically on $h^*$. Indeed, these points are common
roots of $p$ and $F_{h^*}$, restricted to the plane $\pi_{h^*}$.
By B\'ezout's theorem~\cite{Shaf}, either these restricted 
polynomials have a common factor, or they have at most $2d$ roots. 
Since $F_{h^*}$ is irreducible, it must divide the restricted 
$p$, so $p$ must vanish identically on $h^*$, as claimed.

\paragraph{Critical points and parabolas.}
A point $a$ is {\em critical} (or {\em singular})
for a trivariate polynomial $p$ 
if $p(a)=0$ and $\nabla p(a) = 0$; any other point $a$ in 
the zero set of $p$ is called {\em regular}.
A parabola $h^*$ is {\em critical} if all its points are critical.

The following proposition is adapted from \cite{EKS}.
\begin{proposition} \label{factor}
Let $f(x,y,z)$ and $g(x,y,z)$ be two trivariate polynomials,
of respective degrees $k$ and $m$, so that there are 
$km+1$ parabolas of $C$ on which both $f$ and $g$ vanish 
identically.  Then $f$ and $g$ have a common factor.
\end{proposition} 

\noindent{\bf Proof.}
Assume that both $f(x,y,z)$ and $g(x,y,z)$ have a positive degree 
in $x$; this can always be enforced by an appropriate rotation of 
the coordinate frame. It is then an easy exercise to show that
$f$ and $g$ have a common factor if and only if their resultant, 
when viewing them as polynomials in $x$, is identically $0$.
Recall that the resultant is a polynomial in $y$ and $z$.
(The same holds when $f$ and $g$ have any number of variables,
including $x$, in which case the resultant is a polynomial in the
remaining variables.)

For any fixed value $z_0$ of $z$, 
$f(x,y,z_0)$ and $g(x,y,z_0)$ have at least $km+1$ common roots
(at the intersection points of the $km+1$ parabolas with $z=z_0$),
so, by B\'ezout's Theorem \cite{Shaf}, 
they have a common factor. Therefore, 
the resultant, with respect to $x$, of $f(x,y,z_0)$ and 
$g(x,y,z_0)$ is identically $0$ (as a polynomial in $y$). 
Since this is true for every value $z_0$ of $z$, it follows 
that the resultant of $f(x,y,z)$ and $g(x,y,z)$, with respect to $x$, 
vanishes identically as a polynomial in $y$ and $z$. 
Therefore, $f(x,y,z)$ and $g(x,y,z)$, as trivariate polynomials, 
have a common factor.
$\Box$

\begin{proposition} \label{prop1}
Let $C$ be as above. Then
any trivariate square-free polynomial $p$ of degree $d$ can have at 
most $d(d-1)$ critical parabolas in $C$.
\end{proposition} 

\noindent{\bf Proof:}
We prove the claim by induction on the degree $d$ of $p$. The claim
holds trivially for $d=1$, so assume that $d>1$.

Assume first that $p$ is irreducible.
Apply Proposition~\ref{factor} to $p$ and $p_x$, say.
Both polynomials vanish identically on each critical parabola, and their
respective degrees are $d$ and $d-1$. If $p$ had more than 
$d(d-1)$ critical parabolas then $p$ and $p_x$ would have a common
factor, which is impossible since $p$ is irreducible.

Suppose next that $p$ is reducible (but square-free), and write
$p=fg$, so that $f$ and $g$ are non-constant polynomials which
have no common factor (since $p$ is
square-free, this can always be done). Denote the degrees of $f$ and 
$g$ by $d_f$ and $d_g$, respectively; we have $d=d_f+d_g$.

Let $h^*$ be a critical parabola for $p$. Then either $f\equiv 0$ on
$h^*$ or $g\equiv 0$ on $h^*$ (or both). 
Moreover, since $\nabla p = f\nabla g + g\nabla f \equiv 0$ on 
$h^*$, it is easily checked that $h^*$ must satisfy (at least) 
one of the following properties:

\noindent
(i) $f\equiv g\equiv 0$ on $h^*$.

\noindent
(ii) $h^*$ is a critical parabola of $f$.

\noindent
(iii) $h^*$ is a critical parabola of $g$.

Indeed, if (i) does not hold, we have, without loss of generality,
$f\equiv 0$ on $h^*$, but $g$ vanishes only at finitely many points
of $h^*$. On any other point $a$ of $h^*$ we then must have 
$\nabla f(a) = 0$, which implies that $\nabla f$ is identically zero
on $h^*$, so $h^*$ is critical for $f$. This implies (ii); (iii)
holds in the symmetric case where $g\equiv 0$ on $h^*$ but $f$ does
not vanish identically on $h^*$.

By the induction hypothesis, the number of critical parabolas for $f$ is
at most $d_f(d_f-1)$, and the number of critical parabolas for $g$ is
at most $d_g(d_g-1)$. Consider the parabolas that satisfy (i) and
intersect all of them by any of the planes $z=z_0$,
as in the proof of Proposition~\ref{factor}. 
All the intersection points are roots of $f=0$ and $g=0$ on this
plane, and, as follows from the proof of Proposition~\ref{factor},
these bivariate polynomials have no common factor (or, more 
precisely, they can have a common factor only at finitely many 
values of $z$). Hence, 
by B\'ezout's theorem, they have at most $d_fd_g$ common roots. 
Altogether, the number of critical parabolas for $p$ is at most 
$$
d_f(d_f-1) + d_g(d_g-1) + d_fd_g < d(d-1) .
$$
$\Box$

\begin{proposition} \label{prop2}
Let $a$ be a regular point of $p$, so that $p\equiv 0$ on three
parabolas of $C$ passing through $a$. Then these parabolas must have
coplanar tangents at $a$.
\end{proposition} 

\noindent{\bf Proof:}
Any such tangent line must be contained in the tangent plane 
to $p=0$ at $a$.
$\Box$

Hence, a point $a$ incident to three parabolas of $C$ whose
tangent lines at $a$ are non-coplanar, so that $p\equiv 0$ on 
each of these parabolas, must be a critical point of $p$.

\begin{proposition} \label{prop4}
Given a set $S$ of $m$ points in 3-space, there exists a 
nontrivial trivariate polynomial $p(x,y,z)$ which vanishes at all the 
points of $S$, of degree $d$, for any $d$ satisfying 
$\binom{d+3}{3} > m$.
\end{proposition} 

\noindent{\bf Proof:}
(See \cite{EKS,GK}.)
A trivariate polynomial of degree $d$ has 
$\binom{d+3}{3}$ monomials, and requiring it to vanish at 
$m$ points yields these many homogeneous equations in the
coefficients of these monomials. Such an underdetermined system 
always has a nontrivial solution.
$\Box$

\paragraph{Flat points and parabolas.}
Call a regular point $\tau$ of a trivariate polynomial $p$ 
{\em geometrically flat} if it is incident to three distinct 
parabolas of $C$ (with necessarily coplanar tangent lines at $\tau$,
no pair of which are collinear)
on which $p$ vanishes identically.\footnote{%
  Compare this definition with the one in \cite{EKS} (see
  also~\cite{GK}), where a geometrically flat point was defined there
  as a point incident to at least three vanishing {\em lines}, all coplanar.}

Let $\tau$ be a geometrically flat point of $p$, and let 
$h^*_1,h^*_2,h^*_3\in C$ be three incident parabolas on which 
$p$ vanishes. Let $\ttt_i$ denote the tangent line to $h^*_i$ at
$\tau$, and let $v_i$ denote a unit vector in the direction of 
$\ttt_i$, for $i=1,2,3$. 

The second-order Taylor expansion of $p$ at $\tau$ has the form
$$
q(\tau+w) = p(\tau) + \nabla p(\tau)\cdot w + \frac12 w^TH_p(\tau)w =
\nabla p(\tau)\cdot w + \frac12 w^TH_p(\tau)w ,
$$
for any vector $w$, where 
$$
H_p(\tau) = \left(
\begin{array}{ccc}
p_{xx} & p_{xy} & p_{xz} \\
p_{xy} & p_{yy} & p_{yz} \\
p_{xz} & p_{yz} & p_{zz} \\
\end{array}
\right) 
$$
is the {\em Hessian} matrix of $p$.
$q$ is a quadratic polynomial (in $w$) which approximates 
$p$ up to third order terms for sufficiently small values of $|w|$. 

Our goal is to construct, using this approximation and the fact that
$p\equiv 0$ on three parabolas incident to $\tau$, as above, a new
polynomial, depending on $p$, which vanishes at $\tau$, and use this
vanishing as a characterization of flat points.
To do so, we need to make the analysis more specific, 
and taylor it to the special form of $h$-parabolas.

Let $\tau$ be a flat point, and let $a,b,u,v$ be the corresponding 
parameters in the $xy$-plane (so $\tau$ maps $a+tu$ to $b+tv$ for 
each $t\in\reals$; cf.~Remark (1) at the end of (H9)). 
Let $\Sigma=\Sigma(a,b;u,v)$ be the corresponding special surface 
spanned by the parabolas $h^*_{a+tu,b+tv}$, for all $t$
(here we vary $t$ continuously, but only finitely many 
corresponding parabolas belong to $C$).
Since $\tau$ is flat, there exist at least three parabolas
$h^*_{a+t_iu,b+t_iv}$, $i=1,2,3$ (all belonging to $C$,
contained in $\Sigma$, and passing through $\tau$), 
such that $p\equiv 0$ on each of them.

Let $q$ denote, as above, the quadratic polynomial which is the 
second-order Taylor expansion of $p$ at $\tau$. 
Let $h^*=h^*_{a+tu,b+tv}$ be one of the above parabolas on
which $p$ vanishes identically. For $\tau'$ in the vicinity of
$\tau$, $p(\tau')-q(\tau') = O(|\tau'-\tau|^3)$, so, for points 
$\tau'$ near $\tau$ on $h^*$, we have 
$q(\tau') = O(|\tau'-\tau|^3)$. 

Let us continue to consider only points $\tau'$ on $h^*$.
Let $(X_0,Y_0,Z_0)$ (resp., $(X,Y,Z)$) be the coordinates of $\tau$ 
(resp., $\tau'$).  The equations of $h^*$ (see (\ref{parabola})) are
\begin{eqnarray*}
X & = & (a_1+b_1+tu_1+tv_1)Z^2 + 2(a_2+tu_2)Z + (b_1-a_1+tv_1-tu_1) \\
Y & = & (a_2+b_2+tu_2+tv_2)Z^2 - 2(a_1+tu_1)Z + (b_2-a_2+tv_2-tu_2) , 
\end{eqnarray*}
so we have
\begin{eqnarray*}
X - X_0 & = & (Z-Z_0) \left(
(a_1+b_1+tu_1+tv_1)(Z+Z_0) + 2(a_2+tu_2) \right) \\
Y - Y_0 & = & (Z-Z_0) \left(
(a_2+b_2+tu_2+tv_2)(Z+Z_0) - 2(a_1+tu_1) \right) ,
\end{eqnarray*}
which we can further rewrite as
\begin{eqnarray*}
X - X_0 & = & 2(Z-Z_0) \left(
(a_1+b_1+tu_1+tv_1)Z_0 + (a_2+tu_2) \right) +
(Z-Z_0)^2 \left( a_1+b_1+tu_1+tv_1 \right) \\
Y - Y_0 & = & 2(Z-Z_0) \left(
(a_2+b_2+tu_2+tv_2)Z_0 - (a_1+tu_1) \right) +
(Z-Z_0)^2 \left( a_2+b_2+tu_2+tv_2 \right) .
\end{eqnarray*}
Let us simplify these equations as
\begin{eqnarray*}
X-X_0 & = & 2(Z-Z_0)A(t) + (Z-Z_0)^2C(t) \\
Y-Y_0 & = & 2(Z-Z_0)B(t) + (Z-Z_0)^2D(t) ,
\end{eqnarray*}
where $A(t)$, $B(t)$, $C(t)$, and $D(t)$ are all linear functions of
$t$.  If we substitute these equations into the equation of $q$, 
assume that $Z$ is very close to $Z_0$, ignore terms which are 
at least cubic in $Z-Z_0$, and use the fact that
$q(\tau')=O(|\tau'-\tau|^3)$ for any $\tau'$ on $h^*$ sufficiently
close to $\tau$, we conclude that both the linear and the quadratic
parts of $q(\tau')$ (in $Z-Z_0$) vanish identically. 
The linear part is
$$
(Z-Z_0)\nabla p(\tau)\cdot (2A(t),2B(t),1) ,
$$
and the quadratic part is
$$
(Z-Z_0)^2 \left(
\nabla p(\tau)\cdot (C(t),D(t),0) +
\frac12 (2A(t),2B(t),1)^TH_p(\tau)(2A(t),2B(t),1) \right) .
$$
Hence we have
\begin{eqnarray*}
\nabla p(\tau)\cdot (2A(t),2B(t),1) & = & 0 , \\
\nabla p(\tau)\cdot (C(t),D(t),0) +
\frac12 (2A(t),2B(t),1)^TH_p(\tau)(2A(t),2B(t),1) & = & 0 .
\end{eqnarray*}
Note that both equations vanish for (at least) three distinct values
of $t$. Since the first equation is linear in $t$ and the second is
quadratic in $t$, all the coefficients of both equations
are identically zero. Let us
restrict ourselves to the coefficient of the linear term in the first
equation and of the quadratic term in the second one. Denote by
$\alpha$ (resp., $\beta$) the coefficient of $t$ in $A(t)$ (resp.,
$B(t)$). Then we have
\begin{eqnarray*}
\alpha p_X(\tau) + \beta p_Y(\tau) & = & 0 \\
\alpha^2 p_{XX}(\tau) + 2\alpha\beta p_{XY}(\tau) + \beta^2 p_{YY}(\tau) 
& = & 0 .
\end{eqnarray*}
It is easily seen that $\alpha$ and $\beta$ cannot both be zero
(assuming a generic coordinate frame in the original $xy$-plane), so,
eliminating them gives
\begin{equation} \label{sofsof}
p_Y^2(\tau) p_{XX}(\tau) - 2p_X(\tau)p_Y(\tau) p_{XY}(\tau) +
p_X^2(\tau) p_{YY}(\tau) = 0 ,
\end{equation}
which is the constraint we were after.

In what follows, we refer to the left-hand side of (\ref{sofsof}) as
$\Pi(p)$. That is,
$$
\Pi(p) = p_Y^2 p_{XX} - 2p_Xp_Y p_{XY} + p_X^2 p_{YY} ,
$$
and this polynomial has to vanish at $\tau$.

We have thus shown:
\begin{proposition} \label{flat}
Let $p$ be a trivariate polynomial. 
If $\tau$ is a regular geometrically flat point of $p$ 
(with respect to three parabolas of $C$) then $\Pi(p)(\tau) = 0$.
\end{proposition} 

\noindent{\bf Remark.}
Note that the left-hand side of (\ref{sofsof}) is one of the
three polynomials $\Pi_i(p)$ used in \cite{EKS} to analyze flat 
points in a 3-dimensional line arrangement. Specifically,
$$
\Pi(p) = (e_3\times\nabla p)^TH_p(e_3\times\nabla p) ,
$$
where $e_3$ is the unit vector in the $z$-direction; the other
two polynomials are defined analogously, using the other two
coordinate vectors $e_1$, $e_2$.
These polynomials form the {\em second fundamental form} of
$p$; see \cite{EKS,GK} for details.

In particular, if the degree of $p$ is $d$ then the degree 
of $\Pi(p)$ is at most $(d-1)+(d-1)+(d-2) = 3d-4$.

In what follows, we call a point $\tau$ {\em flat} for $p$ if 
$\Pi(p)(\tau) = 0$. We will need the following technical lemma.

\begin{lemma} \label{plinxy}
Let $p$ be an irreducible trivariate polynomial, with the properties
that (i) $\Pi(p)(\tau)=0$ at each regular point $\tau$ of $p=0$, and
(ii) $p\equiv 0$ on at least two distinct intersecting
$h$-parabolas of $C$. Then $p$ is a special polynomial.
\end{lemma}

(Note that the converse of the lemma is trivial, because the
second-order derivatives $p_{XX}$, $p_{XY}$, and $p_{YY}$ are all
identically zero for a special polynomial $p$, and because of the way
such polynomials are constructed.)

\noindent{\bf Proof:}
Fix $Z=Z_0$ and consider the restricted bivariate polynomial
$\tilde{p}(X,Y) = p(X,Y,Z_0)$.
Clearly, $\Pi(\tilde{p}) = \Pi(p)$ on the plane $\pi_0:\;Z=Z_0$. Hence
$\Pi(\tilde{p})=0$ at each regular point $\tau\in\pi_0$ of $p=0$,
and thus at each regular point of $\tilde{p}$.
(Note that a regular point of $\tilde{p}$ is also a regular point of
$p$, although the converse need not be true.)
Note also that $\tilde{p}$ is an irreducible polynomial, except
possibly for finitely many values of $Z_0$.

As is well known~\cite{Gra,Pre}, 
the curvature of the plane curve
$\tilde{p}(X,Y)=0$, at a regular point of $\tilde{p}$, is given by
$$
\kappa = 
\frac{\tilde{p}_Y^2\tilde{p}_{XX} -
2\tilde{p}_X\tilde{p}_Y\tilde{p}_{XY} +
\tilde{p}_X^2\tilde{p}_{YY}}{(\tilde{p}_X^2+\tilde{p}_Y^2)^{3/2}} .
$$
Hence this curve has zero curvature at every regular point of
$\tilde{p}$, and thus, being the zero set of an irreducible 
polynomial, it must be a single line.  
In other words, $p$ is linear in $X$ and $Y$ for every fixed $Z$,
except for finitely many values,
implying that its equation is of the form
$p(X,Y,Z) = A(Z)X+B(Z)Y+C(Z)$, where $A(Z)$, $B(Z)$ and $C(Z)$ are
univariate polynomials. We now exploit assumption (ii), denoting
by $\Sigma$ the unique special surface determined by (and 
containing) the two given $h$-parabolas. The analysis in (H12)
then implies that $\Sigma$ coincides with the zero set of $p$,
so $p$ is indeed a special polynomial, as claimed.
$\Box$

Call an $h$-parabola $h^*\in C$ {\em flat} for $p$ if all the 
points of $h^*$ are flat points of $p$ (with the possible exception 
of a discrete subset). Arguing as in the case of critical
points, if $h^*$ contains more than $2(3d-4)$ flat points then 
$h^*$ is a flat parabola.

As in \cite{EKS,GK}, we next show that, in general, trivariate 
polynomials do not have too many flat parabolas. As before, we 
first establish this property for irreducible polynomials, and 
then extend the analysis to more general polynomials.

\begin{proposition} \label{lflat}
Let $p$ be an irreducible trivariate polynomial of degree $d$,
which is not a special polynomial.
Then $p$ can have at most $3d^2-4d$ flat $h$-parabolas of $C$.
\end{proposition} 

\noindent{\bf Proof:}
Suppose to the contrary that there are more than $3d^2-4d$ flat 
$h$-parabolas. As above, restrict $p$ and $\Pi(p)$ to a fixed plane 
$\pi_0$ of the form $Z=Z_0$.
The number of common roots of $p$ and $\Pi(p)$ on 
$\pi_0$ exceeds the product of their degrees. Since this holds for
every $Z_0$, Proposition~\ref{factor} implies that they must have 
a common factor.  Since $p$ is irreducible, $p$ must be a factor 
of $\Pi(p)$.  This implies that all the (regular) points at which 
$p$ vanishes are flat. Hence, by Lemma~\ref{plinxy}, $p$ must be 
a special polynomial, a contradiction which completes the proof 
of the asserted bound.
$\Box$

\begin{proposition} \label{lflatx}
Let $p$ be any trivariate square-free polynomial of degree $d$ 
with no special polynomial factors. Then $p$ can have at most 
$d(3d-4)$ flat $h$-parabolas in $C$.
\end{proposition} 

\noindent{\bf Proof:}
If $p$ is irreducible, the claim holds by Proposition~\ref{lflat}.
Otherwise, write $p=fg$ where $f$ and $g$ are non-constant 
polynomials with no common factors (and no special polynomial 
factors).
Let $d_f$ and $d_g$ denote their respective degrees, so $d=d_f+d_g$.

Let $\tau$ be a regular flat point of $p$. Then either $f(\tau)=g(\tau)=0$, or
only exactly one of $f(\tau)$, $g(\tau)$ vanishes. Hence, if $h^*$ is a flat
parabola for $p$ then either both $f$ and $g$ vanish identically on
$h^*$ or exactly one of them vanishes identically on $h^*$, while the
other has only finitely many zeroes on $h^*$.

Now, as already argued in the proof of Proposition~\ref{prop1},
there are at most $d_fd_g$
parabolas of the former kind. To handle parabolas of the latter kind,
consider a regular point $\tau$ of $p$ at which $f=0$ but $g$ is  
nonzero.  A simple calculation yields:
\begin{eqnarray*}
p_X & = & f_Xg + fg_X \\
p_Y & = & f_Yg + fg_Y \\
p_{XX} & = & f_{XX}g + 2f_Xg_X + fg_{XX} \\
p_{XY} & = & f_{XY}g + f_Xg_Y + f_Yg_X + fg_{YY} \\
p_{YY} & = & f_{YY}g + 2f_Yg_Y + fg_{YY} .
\end{eqnarray*}
Hence, at $\tau$ we have
\begin{eqnarray*}
p_X(\tau) & = & f_X(\tau)g(\tau) \\
p_Y(\tau) & = & f_Y(\tau)g(\tau) \\
p_{XX}(\tau) & = & f_{XX}(\tau)g(\tau) + 2f_X(\tau)g_X(\tau)  \\
p_{XY}(\tau) & = & f_{XY}(\tau)g(\tau) + f_X(\tau)g_Y(\tau) +
f_Y(\tau)g_X(\tau) \\
p_{YY}(\tau) & = & f_{YY}(\tau)g(\tau) + 2f_Y(\tau)g_Y(\tau) ,
\end{eqnarray*}
and therefore we have at $\tau$, as is easily checked,
$$
\Pi(p)(\tau) = g^3(\tau)\Pi(f)(\tau) .
$$
That is, a regular flat point for $p$, at which $f=0$ but $g$ is
nonzero, is a regular flat point for $f$, and a symmetric statement
holds when $g=0$ but $f$ is nonzero. Hence, any flat parabola of the
latter kind is either a flat parabola for $f$ or a flat parabola for
$g$. Arguing by induction on the degree, the number of flat parabolas
for $p$ is thus at most
$$
3d_f^2-4d_f + 3d_g^2-4d_g + d_fd_g < 3d^2-4d ,
$$
and the lemma follows.
$\Box$

\section{Joint and flat rotations in a set of $h$-parabolas in $\reals^3$}
\label{sec:joints}

In this section we extend the recent algebraic machinery of 
Guth and Katz~\cite{GK}, as further developed by Elekes et
al.~\cite{EKS}, using the algebraic tools set forth in the preceding 
section, to establish the bound $O(n^{3/2})=O(s^3)$
on the number of rotations with multiplicity at least $3$
in a collection of $n$ $h$-parabolas.

\begin{theorem} \label{thm:gk1}
Let $C$ be a set of at most $n$ $h$-parabolas in $\reals^3$, 
and let $P$ be a set of $m$ rotations, each of which is 
incident to at least three parabolas of $C$. 
Suppose further that no special surface contains more than 
$q$ parabolas of $C$. Then $m=O(n^{3/2}+nq)$.
\end{theorem}

\noindent{\bf Remarks.}
{\bf (1)} The recent results of \cite{KSS,Qu} imply that the 
number of joints in
a set of $n$ $h$-parabolas is $O(n^{3/2})$. The proofs in
\cite{KSS,Qu} are much simpler than the proof given below, but they do
not apply to flat points (rotations) as does Theorem~\ref{thm:gk1}. 
Since flat rotations are an integral part of the setup considered in
this paper, we need to count them too, using the stronger
Theorem~\ref{thm:gk1}. Moreover, even if we were to consider only
joint rotations, the analysis of their incidences with the
$h$-parabolas will turn some of them into flat rotations (by 
pruning some of the parabolas), so, as in \cite{EKS}, we will 
need to face flat rotations, no matter what. 

\noindent
{\bf (2)} By Lemma~\ref{sspec}, we always have $q\le s$, and we also have
$n^{1/2} \le s$, so the ``worst-case'' bound on $m$ is $O(ns)$.

\noindent
{\bf (3)} Note that the parameter $n$ in the statement of the theorem is
arbitrary, not necessarily the maximum number $s^2$.
When $n$ attains its maximum possible value $s^2$, the 
bound becomes $m=O(n^{3/2})=O(s^3)$.

The proof of Theorem \ref{thm:gk1} uses the proof technique 
of \cite{EKS}, properly adapted to the present, somewhat more 
involved context of $h$-parabolas and rotations.

\noindent{\bf Proof.} 
We first prove the theorem under the additional assumption that 
$q = n^{1/2}$.
The proof proceeds by induction on $n$, and shows that
$m \le An^{3/2}$, where $A$ is a sufficiently large constant
whose choice will be dictated by the forthcoming analysis.
The statement holds for all $n\le n_0$, for some constant $n_0$, 
if we choose $A$ to be sufficiently large. Fix $n>n_0$, and 
suppose that the claim holds for all $n'<n$.
Let $C$ and $P$ be as in the statement of the theorem, with 
$|C|=n$, and suppose to the contrary that $|P| > An^{3/2}$.

We first apply the following iterative pruning process to $C$.
As long as there exists a parabola $h^*\in C$ incident to fewer 
than $cn^{1/2}$ rotations of $P$, for some constant $1\le c\ll A$ that 
we will fix later, we remove $h^*$ from $C$, remove its 
incident rotations from $P$, and repeat this step with respect 
to the reduced set of rotations.  In this process we delete at 
most $cn^{3/2}$ rotations.  We are thus left with a subset of 
at least $(A-c)n^{3/2}$ of
the original parabolas, each incident to at least 
$cn^{1/2}$ surviving rotations, and each surviving rotation is 
incident to at least three surviving parabolas.
For simplicity, continue to denote these sets as $C$ and $P$.

Choose a random sample $C^s$ of parabolas from $C$, by picking each 
parabola independently with probability $t$, where $t$ is 
a small constant that we will fix later.  

The expected number of parabolas that we choose is $tn_1 \le tn$,
where $n_1$ is the number of parabolas remaining after the pruning.
We have $n_1 = \Omega(n^{1/2})$, because each surviving parabola is 
incident to at least $cn^{1/2}$ surviving rotations, each incident 
to at least two other surviving parabolas; since all these parabolas 
are distinct (recall that a pair of parabolas can meet in at most one 
rotation point), we have $n_1 \ge 2cn^{1/2}$.
Hence, using Chernoff's bound, as in \cite{EKS}
(see, e.g., \cite{AS}), we obtain that, with positive probability, 
(a) $|C^s| \le 2tn$.
(b) Each parabola $h^*\in C$ contains at least $\frac12 ctn^{1/2}$
rotations that lie on parabolas of $C^s$.
(To see (b), take a parabola $h^*\in C$ and a rotation 
$\tau\in P\cap h^*$. Note that $\tau$ will be incident to a parabola 
of $C^s$ with probability at least $t$, so the expected number 
of rotations in $P\cap{h^*}$ which lie on parabolas of $C^s$ is 
at least $ctn^{1/2}$. This, combined with Chernoff's bound, 
implies (b).)

We assume that $C^s$ does indeed satisfy (a) and (b), and
then (recalling that $c\ge 1$)
choose $n^{1/2}$ arbitrary rotations on each parabola
in $C^s$, to obtain a set $S$ of at most $2tn^{3/2}$ rotations.

Applying Proposition~\ref{prop4}, we obtain a nontrivial trivariate 
polynomial $p(X,Y,Z)$ which vanishes at all the rotations of $S$, 
whose degree is at most the smallest integer $d$ satisfying 
$\binom{d+3}{3} \ge |S| + 1$, so
$$
d \le
\lceil (6|S|)^{1/3} \rceil \le (12t)^{1/3}n^{1/2} + 1
\le  2(12t)^{1/3}n^{1/2} ,
$$
for $n$ (i.e., $n_0$) sufficiently large.
Without loss of generality, we may assume
that $p$ is square-free---by removing repeated factors, we get a
square-free polynomial which vanishes on the same set as the 
original $p$, with the same upper bound on its degree.

The polynomial $p$ vanishes on $n^{1/2}$ points on each parabola in $C^s$. 
This number is larger than $2d$, if we choose $t$ sufficiently small so
as to satisfy $4(12t)^{1/3} < 1$. Hence $p$ vanishes identically 
on all these parabolas. Any other parabola of $C$ meets at least 
$\frac12 ctn^{1/2}$ parabolas of $C^s$, at distinct points,
and we can make this number 
also larger than $2d$, with an appropriate choice of $t$ and $c$ 
(we need to ensure that $ct > 8(12t)^{1/3}$). 
Hence, $p$ vanishes identically on each parabola of $C$.

We will also later need the property that each parabola of $C$ 
contains at least $9d$ points of $P$; that is, we require that 
$cn^{1/2} > 9d$, which will hold if $c>18(12t)^{1/3}$.

To recap, the preceding paragraphs impose several inequalities on $c$
and $t$, and a couple of additional similar inequalities will be 
imposed later on. All these inequalities are easy to satisfy by 
choosing $t<1$ to be a sufficiently small positive constant, and 
$c$ a sufficiently large constant. (These choices will also
affect the choice of $A$---see below.)

We note that $p$ can have at most $d/3$ special polynomial factors
(since each of them is a cubic polynomial); i.e., $p$ can vanish 
identically on at most $d/3$ respective special surfaces 
$\Xi_1,\ldots,\Xi_k$, for $k\le d/3$.  We factor out all these
special polynomial factors from $p$, and let $\tilde{p}$ denote 
the resulting polynomial, which is a square-free polynomial 
without any special polynomial factors, of degree at most $d$.

Consider one of the special surfaces $\Xi_i$, and let $t_i$ 
denote the number of parabolas contained in $\Xi_i$. 
Then any rotation on $\Xi_i$ is either an intersection point of 
(at least) two of these parabolas, or it lies on at most one of 
them. The number of rotations of the first kind is $O(t_i^2)$. 
Any rotation $\tau$ of the second kind is incident to at least 
one parabola of $C$ which crosses $\Xi_i$ transversally at $\tau$. 
We note that each $h$-parabola $h^*$ can cross $\Xi_i$ in at
most three points. Indeed, substituting the equations of $h^*$ 
into the equation $E_2(Z)X-E_1(Z)Y+K(Z)=0$ of $\Xi_i$ 
(see (\ref{sigma})) yields a cubic equation in $Z$, 
with at most three roots. Hence, the number of rotations 
of the second kind is $O(n)$, and the overall number of 
rotations on $\Xi_i$ is $O(t_i^2+n) = O(n)$, since we have
assumed in the present version of the proof that $t_i \le n^{1/2}$.

Summing the bounds over all surfaces $\Xi_i$, we conclude that
altogether they contain $O(nd)$ rotations, which we bound by 
$bn^{3/2}$, for some absolute constant $b$.

We remove all these vanishing special surfaces, together with 
the rotations and the parabolas which are fully contained in them, 
and let $C_1\subseteq C$ and $P_1\subseteq P$ denote, respectively,
the set of those parabolas of $C$ (rotations of $P$) which are not
contained in any of the vanishing surfaces $\Xi_i$. 

Note that there are still at least three parabolas of $C_1$
incident to any remaining rotation in $P_1$, since none of the 
rotations of $P_1$ lie in any surface $\Xi_i$, so all parabolas 
incident to such a rotation are still in $C_1$.

Clearly, $\tilde{p}$ vanishes identically on every $h^*\in C_1$. 
Furthermore, every $h^* \in C_1$ contains at most $d$ points 
in the surfaces $\Xi_i$, because, as just argued, it crosses each
surface $\Xi_i$ in at most three points.

Note that this also holds for every parabola $h^*$
in $C\setminus C_1$, if we only count intersections of $h^*$ with 
surfaces $\Xi_i$ which do not fully contain $h^*$.

Hence, each $h^*\in C_1$ contains at least $8d$ rotations of $P_1$. 
Since each of these rotations is incident to at least three 
parabolas in $C_1$, each of these rotations is either critical 
or geometrically flat for $\tilde{p}$.

Consider a parabola $h^* \in C_1$. If $h^*$ contains more than 
$2d$ critical rotations then $h^*$ is a critical parabola for 
$\tilde{p}$. By Proposition~\ref{prop1}, the number of such parabolas
is at most $d(d-1)$. Any other parabola $h^*\in C_1$ contains more 
than $6d$ geometrically flat points and hence $h^*$ must be a flat 
parabola for $\tilde{p}$. By Proposition \ref{lflatx}, the number 
of such parabolas is at most $d(3d-4)$. Summing up we obtain
$$
|C_1| \le d(d-1)+d(3d-4) < 4d^2 .
$$
We require that $4d^2 < n/2$; that is, 
$32(12t)^{2/3} < 1$, which can be guaranteed by choosing $t$
sufficiently small. 

We next want to apply the induction hypothesis to $C_1$, with the
parameter $4d^2$ (which dominates the size of $C_1$).
For this, we first need to argue that each special
surface contains at most $(4d^2)^{1/2} = 2d$ parabolas of $C_1$.
Indeed, let $\Xi$ be a special surface.
Using (\ref{sigma}), eliminate, say, $Y$ from the equation
of $\Xi$ and substitute the resulting expression into the equation
of $\tilde{p}$, to obtain a bivariate polynomial $\tilde{p}_0(X,Z)$.
Let $h^*$ be a parabola of $C_1$ contained in $\Xi$. We represent
$h^*$ by its $X$-equation of the form $X=Q(Z)$, and observe that
$\tilde{p}_0(X,Z)$ vanishes on the zero set of $X-Q(Z)$. 
Hence $\tilde{p}_0$ must be divisible by $X-Q(Z)$. Note that, in a
generic coordinate frame in the $xy$-plane, two different parabolas 
cannot have the same equation $X=Q(Z)$, because this equation uniquely
determines $a_1,b_1$, and $a_2$, and then, in a generic frame, $b_2$
is also uniquely determined. Note also that the degree of $\tilde{p}_0$
is at most $3d$, and that the degree of each factor $X-Q(Z)$ is $2$, 
implying that $\Sigma$ can contain at most $3d/2$ parabolas of $C_1$.

An important observation, which we will use in the proof of general
version of the theorem, is that the argument just given does not 
use the assumed bound on the number of $h$-parabolas contained in 
a special surface, but, rather, establishes this bound ``from
scratch'' for the 
subproblem involving $P_1$ and $C_1$. That is, even if the original 
problem does not satisfy the extra assumption in the restricted
version, the subproblems that it generates always do satisfy it.

Hence, the maximum number of parabolas of $C_1$ contained in a 
special surface is at most $3d/2\le (4d^2)^{1/2}$, so, by the 
induction hypothesis, the number of points in $P_1$ is at most 
$$
A(4d^2)^{3/2} \le \frac{A}{2^{3/2}} n^{3/2} .
$$
Adding up the bounds on the number of points on parabolas removed 
during the pruning process and on the special surfaces $\Xi_i$ 
(which correspond to the special polynomial factors of $p$), 
we obtain
$$
|P| \le 
\frac{A}{2^{3/2}}n^{3/2} + (b+c)n^{3/2} \le An^{3/2} \ ,
$$
with an appropriate, final choice of $t$, $c$, and $A$. 
This contradicts the assumption that $|P| > A n^{3/2}$, and thus
establishes the induction step for $n$, and, consequently,
completes the proof of the restricted version of the theorem.

\paragraph{Proof of the general version:}
The proof proceeds almost exactly as the proof of the restricted
version, except for the analysis of the number of rotations on 
the special surfaces $\Xi_i$.  As noted above, we encounter this 
difference only once, in handling the original problem: When we 
apply the induction step, we always fall into the restricted 
setup.


By assumption, each special surface $\Xi_i$ contains at most $q$ 
$h$-parabolas. We modify the preceding analysis, so that each
parabola is considered only once. That is, we iterate over the special
surfaces in some order. When handling a surface $\Xi_i$, we consider
only those $h$-parabolas that are not contained in any previously 
processed surface, and bound the number of rotations that they 
contain. Then we remove these parabolas and rotations from further 
considerations and go to the next surface.

As argued above, a special surface $\Xi_i$ containing $t_i$
(surviving) parabolas contains at most $O(t_i^2+n)$ rotations which
lie on these parabolas (and on no previously processed parabola).
Summing these bounds over all special surfaces, and using the fact
that $t_i\le q$ for each $i$, we get an overall bound
$O(nd + q\sum_i t_i) = O(n^{3/2} + nq)$, as asserted.
$\Box$

We summarize the remarks following Theorem~\ref{thm:gk1}, combined
with Lemma~\ref{lem:lower}, in the following corollary.

\begin{corollary} \label{roth}
Let $S$ be a set of $s$ points in the plane. Then there are at most
$O(s^3)$ rotations which map some (degenerate or non-degenerate)
triangle spanned by $S$ to another (congruent and equally 
oriented) such triangle. This bound is tight in the worst case.
\end{corollary}
 
In the following section we will continue to adapt the analysis of
\cite{EKS} to obtain bounds on the number of incidences between
helices ($h$-parabolas) and rotations with multiplicity $\ge 3$, 
and, consequently, obtain bounds on $|P_{\ge k}|$, for any $k\ge 3$.

\section{Incidences between parabolas and rotations}
\label{sec:inc}

In this section we further adapt the machinery of \cite{EKS} to derive 
an upper bound on the number of incidences between $m$ rotations 
and $n$ $h$-parabolas in $\reals^3$, where each rotation is
incident to at least three parabolas (i.e., has multiplicity $\ge 3$).

\begin{theorem} \label{inc-gen}
For an underlying ground set $S$ of $s$ points in the plane, let
$C$ be a set of at most $n\le s^2$ $h$-parabolas defined on $S$,
and let $P$ be a set of $m$ rotations with multiplicity at least $3$
(with respect to $S$). Then
$$
I(P,C) = O\left(m^{1/3}n + m^{2/3}n^{1/3}s^{1/3}\right) .
$$
\end{theorem}

\noindent{\bf Remark.}
As easily checked, the first term dominates the second term
when $m\le n^2/s$, and the second term dominates when 
$n^2/s < m \le ns$. In particular, the first term dominates when
$n=s^2$, because we have $m=O(s^3)=O(n^2/s)$

\noindent{\bf Proof:}
The proof of Theorem~\ref{inc-gen} proceeds in two steps. We first
establish a bound which is independent of $m$, and then apply it to
obtain the $m$-dependent bound asserted in the theorem. 

For the first step, we have:
\begin{theorem} \label{inc-32}
Let $C$ be a set of at most $n\le s^2$ $h$-parabolas defined on $S$,
and let $P$ be a set of rotations with multiplicity at least $3$
with respect to $S$, such that no special surface contains more 
than $n^{1/2}$ parabolas of $C$. Then the number of incidences 
between $P$ and $C$ is $O(n^{3/2})$.
\end{theorem}

\noindent{\bf Proof.}
Write $I=I(P,C)$ for short, and put $m=|P|$.
We will establish the upper bound $I\le Bn^{3/2}$, for some
sufficiently large absolute constant $B$, whose specific choice will
be dictated by the various steps of the proof. Suppose then to the
contrary that $I>Bn^{3/2}$ for the given $C$ and $P$.

For $h^*\in C$, let $\nu(h^*)$ denote the number of rotations 
incident to $h^*$. We refer to $\nu(h^*)$ as the 
{\em multiplicity} of $h^*$.
We have $\sum_{h^*\in C} \nu(h^*) = I$.
The average multiplicity of a parabola $h^*$ is $I/n$.

We begin by applying the following pruning process.  Put
$\nu = I/(6n)$. As long as there exists a parabola $h^*\in C$ 
whose multiplicity is smaller than $\nu$, we remove $h^*$ 
from $C$, but do not remove any rotation incident to $h^*$. 
We keep repeating this step (without changing $\nu$), until 
each of the surviving parabolas has multiplicity at least $\nu$. 
Moreover, if, during the pruning process, some rotation $\tau$ loses 
$\lfloor \mu(\tau)/2\rfloor$ incident parabolas, we remove $\tau$ from $P$. 
This decreases the multiplicity of some parabolas, and we use the 
new multiplicities in the test for pruning further parabolas, but 
we keep using the original threshold $\nu$.

When we delete a parabola $h^*$, we lose at most $\nu$ 
incidences with surviving rotations. When a rotation $\tau$ is removed, 
the number of current incidences with $\tau$ is smaller than or 
equal to twice the number of incidences with $\tau$ that have 
already been removed. Hence, the total number of incidences 
that were lost during the pruning process is a most $3n\nu = I/2$.
Thus, we are left with a subset $P_1$ of the rotations and with a
subset $C_1$ of the parabolas, so that each $h^*\in C_1$ is 
incident to at least $\nu=I/(6n)$ rotations of $P_1$, and each 
rotation $\tau\in P_1$ is incident to at least three parabolas of $C_1$ 
(the latter is an immediate consequence of the rule for pruning 
a rotation).  Moreover, we have $I(P_1,C_1) \ge I/2$. 
It therefore suffices to bound $I(P_1,C_1)$.

Let $n_1=|C_1|$. Since at least three parabolas in $C_1$ are 
incident to each rotation in $P_1$, it follows that each parabola 
in $C_1$ is incident to at most $n_1/2$ rotations of $P_1$, and 
therefore $I(P_1,C_1) \le n_1^2/2$. Combining this with the 
fact that $I(P_1,C_1) \ge I/2$, we get that 
$n_1 \ge B^{1/2}n^{3/4}$.

We fix the following parameters
$$
x = \frac{n_1}{n^{1/2}}
\quad\quad\mbox{and}\quad\quad
t = \delta \frac{n_1}{n} ,
$$
for an appropriate absolute constant $\delta < 1$, whose value will be
fixed shortly. Clearly, $t < 1$, and we can also ensure that $x < \nu$,
i.e., that $I > 6 n_1 n^{1/2}$, by choosing $B > 6$. Furthermore,
since $n_1 \ge B^{1/2}n^{3/4}$ we have $x \ge B^{1/2}n^{1/4}$.

We construct a random sample $C^s_1$ of parabolas of $C_1$ by choosing
each parabola independently at random with probability $t$; the expected
size of $C^s_1$ is $tn_1$. Now take $x$ (arbitrary) rotations of $P_1$ 
on each parabola of $C^s_1$ (which can always be done since $x<\nu$),
to form a sample $S$ of rotations in $P_1$, of expected size at most
$txn_1$. 

For any parabola $h^*\in C_1$, the expected number of rotations of
$P_1\cap{h^*}$ which lie on parabolas of $C^s_1$ is at least $t\nu$
(each of the at least $\nu$ rotations $a\in P_1\cap{h^*}$ is incident 
to at least one other parabola of $C_1$, and the probability of this 
parabola to be chosen in $C^s_1$ is $t$).
We assume that $B$ is large enough so that
${\displaystyle t\nu = \delta\frac{n_1}{n}\frac{I}{6n} \ge 
\frac{\delta B}{6} \frac{n_1}{n^{1/2}} }$ is larger than $2x$
(it suffices to choose $B > 12/\delta$).
Since $t\nu> 2x  = \Omega({n^{1/4}})$, and the expected size
of $C_1^s$ is ${\displaystyle tn_1 = \frac{\delta n_1^2}{n} \ge
B\delta n^{1/2} }$, we can use Chernoff's bound, to show
that there exists a sample $C^s_1$ such that
(i) $|C^s_1| \le 2tn_1$, and
(ii) each parabola $h^*\in C_1$ contains at least $\frac12 t\nu > x$
rotations of $P_1$ which lie on parabolas of $C^s_1$.
In what follows, we assume that $C^s_1$ satisfies these properties.
In this case, we have $|S|\le 2txn_1$.

Now construct, using Proposition~\ref{prop4}, a nontrivial suqare-free
trivariate polynomial $p$ which vanishes on $S$, of smallest 
degree $d$ satisfying $\binom{d+3}{3} \ge |S|+1$, so
\begin{eqnarray*}
d & \le &
\lceil (6|S|)^{1/3} \rceil \le (12txn_1)^{1/3} + 1 =
 (12\delta)^{1/3} \frac{n_1}{n^{1/2}} + 1   \\
& \le & 2(12\delta)^{1/3} \frac{n_1}{n^{1/2}} 
\end{eqnarray*}
for $n$ sufficiently large (for small values of $n$ we ensure the
bound by choosing $B$ sufficiently large, as before).

We will choose $\delta < 1/6144$, so $x > 4d$.

As above, and without loss of generality, we may assume that $p$ 
is square-free: factoring out repeated factors only lowers the 
degree of $p$ and does not change its zero set.

The following properties hold:
(a) Since $x>2d$, $p$ vanishes at more than $2d$ rotations on each
parabola of $C^s_1$, and therefore, as already argued, it vanishes 
identically on each of these parabolas.
(b) Each parabola $h^*\in C_1$ contains at least 
$\frac12 t\nu > x > 2d$ rotations which lie on parabolas of $C^s_1$. 
Since, as just argued, $p$ vanishes at these rotations, it must 
vanish identically on $h^*$. 
Thus, $p\equiv 0$ on every parabola of $C_1$.

Before proceeding, we enforce the inequality $d^2 < \frac18 n_1$
which will hold if we choose $\delta$ so that $(12\delta)^{2/3} < 1/32$.
Similarly, an appropriate choice of $\delta$ (or $B$) also ensures 
that $\nu > 9d$.

We next consider all the special polynomial factors of $p$, and factor
them out, to obtain a square-free polynomial $\tilde{p}$, of degree at
most $d$, with no special polynomial factors. As in the previous
analysis, $p$ can have at most $d/3$ special polynomial factors, 
so it can vanish identically on at most $d/3$ special surfaces
$\Xi_1,\ldots,\Xi_k$, for $k\le d/3$.  
Let $C_2\subseteq C_1$ denote the set of those parabolas of $C_1$ 
which are not contained in any of the vanishing surfaces $\Xi_i$. 
For each parabola $h^*\in C_2$, $\tilde{p}$ vanishes identically 
on $h^*$, and (as argued above) at most $d$ rotations in 
$P_1\cap{h^*}$ lie in the surfaces $\Xi_i$. Hence, $h^*$
contains at least $8d$ remaining rotations, each of which is either
critical or flat for $\tilde{p}$, because each such point is 
incident to at least three parabolas (necessarily of $C_2$)
on which $\tilde{p}\equiv 0$.

Hence, either at least $2d$ of these rotations are critical, and
then $h^*$ is a critical parabola for $\tilde{p}$, or
at least $6d$ of these rotations are flat, and then $h^*$ is
a flat parabola for $\tilde{p}$. Applying Propositions~\ref{prop1}
and \ref{lflatx}, the overall number of parabolas in $C_2$ is
therefore at most
$$
d(d-1)+d(3d-4) < 4d^2 < \frac12 n_1 .
$$
On the other hand, by assumption, each vanishing special surface 
$\Xi_i$ contains at most $n^{1/2}$ parabolas of $C$, so the number of
parabolas contained in the special vanishing surfaces is at most
$n^{1/2}d < \frac14 n^{1/2}x \le \frac14 n_1$,
with our choice of $\delta$.

Hence, the overall number of parabolas in $C_1$ is smaller than
$\frac12 n_1 + \frac14 n_1 < n_1$, a contradiction 
that completes the proof of Theorem~\ref{inc-32}.
$\Box$

\noindent{\bf Proof of Theorem~\ref{inc-gen}.}
Write $I=I(P,C)$ for short.  
Set $\nu = cm^{1/3}$ and $\mu=cn/m^{2/3}$, for some sufficiently
large constant $c$ whose value will be determined later, and 
apply the following pruning process.
As long as there exists a parabola $h^*\in C$ 
whose multiplicity is smaller than $\nu$, we remove $h^*$ from $C$, 
but do not remove any rotation incident to $h^*$. Similarly, as 
long as there exists a rotation $\tau\in P$ whose multiplicity 
is smaller than $\mu$, we remove $\tau$ from $P$.
Of course, these removals may reduce the multiplicity of some
surviving rotations or parabolas, making additional rotations and
parabolas eligible for removal. We keep repeating this step 
(without changing the initial thresholds $\nu$ and $\mu$), 
until each of the surviving parabolas has multiplicity at 
least $\nu$ and each of the surviving rotations has multiplicity 
at least $\mu$.  We may assume that $\mu\ge 3$, by choosing $c$ 
suficiently large and using Theorem~\ref{thm:gk1}(i).

When we delete a parabola $h^*$, we lose at most $\nu$ incidences 
with surviving rotations. When a rotation $\tau$ is removed, we 
lose at most $\mu$ incidences with surviving parabolas. 
All in all, we lose at most $n\nu + m\mu = 2c m^{1/3}n$ incidences,
and are left with a subset $P_1$ of $P$ and with a
subset $C_1$ of $C$, so that each parabola of $C_1$ is incident
to at least $\nu$ rotations of $P_1$, and each rotation of $P_1$ is
incident to at least $\mu$ parabolas of $C_1$ (these subsets might be
empty). Put $n_1=|C_1|$ and $m_1=|P_1|$.
We have $I \le I(P_1,C_1) + 2c m^{1/3}n$, so it
remains to bound $I(P_1,C_1)$, which we do as follows.

We fix some sufficiently small positive parameter $t<1$, and 
construct a random sample $P_1^s\subset P_1$ by choosing each rotation
of $P_1$ independently with probability $t$. The expected size of
$P_1^s$ is $m_1t$, and the expected number of points of $P_1^s$ on
any parabola of $C_1$ is at least $\nu t = ctm^{1/3}$.
Chernoff's bound implies that, with positive probability, 
$|P_1^s|\le 2m_1t$, and $|P_1^s\cap h^*| \ge \frac12 ctm^{1/3}$ for
every $h^*\in C_1$. We can therefore assume that $P_1^s$ satisfies
all these inequalities. (For the bound to apply, $m_1$ (and $m$)
must be at least some sufficiently large constant; if this is not 
the case, we turn the trivial bound $m_1n$ (or $mn$) on $I$ into 
the bound $O(m_1^{1/3}n)$ (or $O(m^{1/3}n)$) by choosing the 
constant of proportionality sufficiently large.)

Construct, using Proposition~\ref{prop4}, a nontrivial square-free
trivariate polynomial $p$ which vanishes on $P_1^s$, whose degree
is at most the smallest integer $d$ satisfying
${d+3\choose 3} \ge 2tm_1+1$, so
$$
d \le \lceil (12tm_1)^{1/3} \rceil \le 3t^{1/3}m_1^{1/3} ,
$$
assuming (as above) that $m_1$ is sufficiently large.

Choosing $c$ to be large enough, we may assume that $\nu t > 18d$.
(This will hold if we ensure that $ct > 54t^{1/3}$.)
This implies that $p$ vanishes at more than $9d$ points on each 
parabola $h^*\in C_1$, and therefore it vanishes identically on 
each of these parabolas.

As in the previous analysis, we factor out the special polynomial 
factors of $p$, obtaining a square-free polynomial $\tilde{p}$, 
of degree at most $d$, with no special polynomial factors. 
Let $\Xi_1,\ldots,\Xi_k$ denote the special surfaces on which 
$p$ vanishes identically (the zero sets of the special 
polynomial factors of $p$), for some $k\le d/3$.

Let $C_2\subseteq C_1$ (resp., $P_2\subseteq P_1$) 
denote the set of those parabolas of $C_1$ (resp., rotations 
of $P_1$) which are not contained in any of the vanishing 
surfaces $\Xi_i$.
Put $C'_2 = C_1\setminus C_2$ and $P'_2 = P_1\setminus P_2$.

For each parabola $h^*\in C_2$, $\tilde{p}$ vanishes identically 
on $h^*$, and, as argued in the proof of Theorem~\ref{thm:gk1},
at most $d$ rotations of $P_1\cap{h^*}$ lie in
the surfaces $\Xi_i$. Hence, $h^*$ contains more than $8d$ 
rotations of $P_2$, and, arguing as in the preceding proof,
each of these rotations is either critical or flat for $\tilde{p}$.
Hence, either more than $2d$ of these rotations are critical, and
then $h^*$ is a critical parabola for $\tilde{p}$, or more than 
$6d$ of these rotations are flat, and then $h^*$ is a flat 
parabola for $\tilde{p}$. Applying Propositions~\ref{prop1} and 
\ref{lflatx}, the overall number of parabolas in $C_2$ is
therefore at most
$$
d(d-1)+d(3d-4) < 4d^2 .
$$
We now apply Theorem~\ref{inc-32} to $C_2$ and $P_2$, with 
the bound $4d^2$ on the size of $C_2$. The conditions
of this theorem hold for these sets: Clearly, each rotation 
in $P_2$ is incident to at least three parabolas of $C_2$.
For the other condition, we argue exactly as in the proof of
Theorem~\ref{thm:gk1}, to conclude that any special surface
can contain at most $3d/2$ parabolas of $C_1$, establishing the second
condition of Theorem~\ref{inc-32}. This theorem then implies 
that the number of incidences between $P_2$ and $C_2$, which 
is also equal to the number of incidences between $P_2$ and 
$C_1$, is
$$
I(P_2,C_1) = I(P_2,C_2) = O((4d^2)^{3/2}) = O(d^3) = O(m) \ .
$$
Moreover, since each parabola of $C_2$ contains at least eight times
more rotations of $P_2$ than of $P'_2$, this bound also applies to
the number of incidences between $P'_2$ and $C_2$.

It therefore remains to bound the number of incidences between 
$P'_2$ and $C'_2$, namely, between the rotations and parabolas
contained in the vanishing special surfaces $\Xi_i$.
To do so, we iterate over the surfaces, say, in the order
$\Xi_1,\ldots,\Xi_k$. For each surface $\Xi_i$ in turn,
we process the rotations and parabolas contained in $\Xi_i$ and
then remove them from further processing on subsequent surfaces.

Let us then consider a special surface $\Xi_i$.
Let $m_i$ and $n_i$ denote respectively the number of rotations 
and parabolas contained in $\Xi_i$, which were not yet removed 
when processing previous surfaces.
The number of incidences between these rotations and parabolas
can be bounded by the classical Szemer\'edi-Trotter
incidence bound~\cite{ST} (see also (\ref{weak23})), which is
$O(m_i^{2/3}n_i^{2/3} + m_i+n_i)$.
Summing these bounds over all the special surfaces $\Xi_i$, and using
H\"older's inequality and the fact, established in Lemma~\ref{sspec},
that $n_i \le s$, we get an overall bound of
$$
O\left( \sum_i \left( m_i^{2/3}n_i^{2/3} + m_i + n_i \right) \right) = 
$$
$$
O\left( s^{1/3} 
  \sum_i m_i^{2/3}n_i^{1/3} + \sum_i (m_i + n_i) \right) = 
O\left( m^{2/3}n^{1/3}s^{1/3} + m + n \right) ,
$$
where we use the facts that $\sum_i m_i \le m$ and
$\sum_i n_i \le n$, which follow since in this analysis each parabola
and rotation is processed at most once.
The two linear terms satisfy 
$n = O(m^{1/3}n)$ (the bound obtained in the pruning process), and
$m = O(m^{2/3}n^{1/3}s^{1/3})$ since $m=O(ns)$; see Remark (2) 
following Theorem~\ref{thm:gk1}.

We are not done yet, because each rotation of $P'_2$ is processed 
only once, within the first surface $\Xi_i$ containing it. This,
however, can be handled as in \cite{EKS}. That is, let $\tau$ be a
rotation which was processed within the first surface $\Xi_i$
containing it. Suppose that $\tau$ also lies on some later surface
$\Xi_j$, with $j>i$, and let $h^*$ be a parabola contained in $\Xi_j$,
which has not been removed yet; in particular, $h^*$ is not contained
in $\Xi_i$, and thus meets it transversally, so the incidence between
$h^*$ and $\tau$ can be regarded as one of the transversal incidences
in $\Xi_i$, which we have been ignoring so far. 
To count them, we simply recall that each
parabola, whether of $C'_2$ or of $C_2$, has at most three
transversal intersections with a surface $\Xi_i$ 
(see the proof of Theorem~\ref{thm:gk1}), for a total of
at most $d$ crossings with all the vanishing surfaces. Since each of
these parabolas contains at least $9d$ rotations of $P_1$, those
``transversal incidences'' are only a fraction of the total number of
incidences, and we simply ignore them altogether.

To recap, we obtain the following bound on the number of incidences 
between $P_1$ and $C_1$:
$$
I(P_1,C_1) = O\left( m + m^{1/3}n + m^{2/3}n^{1/3}s^{1/3} \right) =
O\left(m^{1/3}n + m^{2/3}n^{1/3}s^{1/3}\right) .
$$
Adding the bound $2c m^{1/3}n$ on the incidences lost during the
pruning process, we get the asserted bound.
$\Box$

It is interesting to note that the proof technique also yields the
following result.

\begin{corollary} \label{mgek}
Let $C$ be a set of $n$ $h$-parabolas and $P$ a set of points in 3-space
which satisfy the conditions of Theorem~\ref{inc-gen}(i). Then, for any
$k\ge 1$, the number $M_{\ge k}$ of points of $P$ incident to at
least $k$ parabolas of $C$ satisfies
$$
M_{\ge k} =
\begin{cases}
{\displaystyle O\left( \frac{ns}{k^{3}} \right) } &
  \mbox{for $k\le s^{2/3}/n^{1/3}$,} \\
{\displaystyle O\left( \frac{n^{3/2}}{k^{3/2}} \right) } &
  \mbox{for $s^{2/3}/n^{1/3}\le k\le n^{1/3}$,} \\
{\displaystyle O\left( \frac{n^2}{k^3} + \frac{n}{k} \right) } &
  \mbox{for $k > n^{1/3}$.}
\end{cases}
$$
\end{corollary}
\noindent{\bf Proof:}
Write $m = M_{\ge k}$ for short.  We clearly have $I(P,C) \ge km$.
Theorem~\ref{inc-gen} then implies 
$km = O(m^{1/3}n+m^{2/3}n^{1/3}s^{1/3})$, from which the first two bounds
follow. If $k > n^{1/3}$ we use the other bound
(in (\ref{weak23})), to obtain $km = O(m^{2/3}n^{2/3}+m+n)$, which
implies that $m = O(n^2/k^3 + n/k)$
(which is in fact an equivalent statement of the classical
Szemer\'edi-Trotter bound).
$\Box$


\section{Further improvements}
\label{sec:impr}

In this section we further improve the bound in Theorem~\ref{inc-gen}
(and Corollary~\ref{mgek}) 
using more standard space decomposition techniques. We show:
\begin{theorem} \label{thm:impr}
The number of incidences between $m$ arbitrary rotations and $n$
$h$-parabolas, defined for a planar ground set with $s$ points, is
$$
O^*\left(m^{5/12}n^{5/6}s^{1/12} +
m^{2/3}n^{1/3}s^{1/3} + n \right) ,
$$
where the $O^*(\cdot)$ notation hides polylogarithmic factors.
In particular, when all $n=s^2$ $h$-parabolas are considered, the
bound is
$$
O^*\left( m^{5/12}s^{7/4} + s^2 \right) .
$$
\end{theorem} 

\noindent{\bf Proof:}
We dualize the problem as follows. We map each parabola $h^*_{a,b}$ 
to the point $\hat{h}_{a,b} = (a,b) = (a_1,a_2,b_1,b_2)$ in $\reals^4$. 
Each rotation $\tau$ is mapped to a 2-plane $\hat{\tau}$, which 
is the locus of all points $\hat{h}$ such that $\tau$ is incident to
$h^*$. This is indeed a 2-plane, because the equations of $\tau$,
either (\ref{eq:helix}) in the $(\xi,\eta,\theta)$-frame, or
(\ref{parabola}) in the $(X,Y,Z)$-frame, are a pair of linear 
(independent) equations in $(a_1,a_2,b_1,b_2)$.

So in this new setup we have $n$ points and $m$ 2-planes in 4-space,
and we wish to bound the number of incidences between these points and
2-planes. We note that any pair of these 2-planes intersect in at most
one point. (The corresponding statement in the primal setup is that two
rotations can be incident to at most one common $h$-parabola.)

To bound the number of incidences, we first project the points and
2-planes onto the 3-space $b_2=0$. We claim that, with a generic choice 
of the coordinate frame in the original $xy$-plane, the projected 
points remain distinct. Indeed, a point $(a_1,a_2,b_1,b_2)$, 
dual to an $h$-parabola $h^*_{a,b}$, is projected to the point
$(a_1,a_2,b_1)$, so the projected point uniquely determines $a$,
and also $b$, because we may assume that no two points of $S$ have 
the same $x$-coordinate $b_1$. Hence the projected points are all 
distinct. 

This is not necessarily the case for the 2-planes. Indeed, consider
a 2-plane $\hat{\tau}$. Its projection onto the $a_1a_2b_1$-space
is the plane satisfying the first equation of (\ref{parabola}), say, 
namely
$$
X = (a_1+b_1)Z^2 + 2a_2Z + (b_1-a_1) .
$$
It is easily checked that this equation uniquely determines the 
$X$ and $Z$ components of $\tau$, leaving $Y$ (i.e., the shift 
along the $y$-direction that $\tau$ makes after its initial
pure rotation) undetermined. Thus it is possible that several 
distinct rotations, all with the same $X$ and $Z$ components,
are projected to the same 2-plane. This has the potential danger 
that the projection loses incidences, when several 2-planes,
incident to a common point $\hat{\tau}$, get projected into the same
plane, so that, instead of several incidences with $\hat{\tau}$ in
4-space, we get only one incidence in the projection. Nevertheless,
this bad situation cannot arise. This follows from the easy
observation that two distinct rotations with the same $X$ and $Z$
components cannot both map a point $(a_1,a_2)$ into the same point
$(b_1,b_2)$.

To recap, after the projection we get $n$ points and at most $m$ 
planes in $\reals^3$, and our goal is to bound the number of 
incidences between them. More precisely, we want to bound only 
the number of original incidences. We note that each such incidence
appears as an incidence in the projection, but not necessarily the
other way around. We recall that, in general, the number of incidences
between $n$ points and $m$ planes in 3-space can be $mn$ in the worst
case, because of the possibility that many points lie on a common line
and many planes pass through that line. This situation can also arise
in our setup, but we will apply a careful analysis to show that the
number of original incidences that project to such a degenerate
configuration is much smaller.

We proceed as follows. We fix a parameter $r$, to be determined 
shortly, and construct the following decomposition of 3-space.
First, we note that the projected points $(a_1,a_2,b_1)$ have 
only $s$ distinct $a_1$-coordinates, which are the $x$-coordinates 
of the points of $S$. Similarly, they have only $s$ distinct
$b_1$-coordinates. We partition the 3-space by a set $R_1$ of $r$ 
planes orthogonal to the $a_1$-axis, so that within 
each resulting slab the projected points have at most $s/r$ 
distinct $a_1$-coordinates. We construct a similar collection
$R_2$ of $r$ planes orthogonal to the $b_1$-axis, so that within 
each resulting slab the projected points have at most $s/r$ 
distinct $b_1$-coordinates. 
We then choose a random sample $R_0$ of $r$ of the projected 
planes. We take the set $R=R_0\cup R_1\cup R_2$ of $3r$ planes, 
construct their arrangement, and decompose each of its cells into
simplices. We obtain $O(r^3)$ simplices, and the construction and 
the standard $\eps$-net theory \cite{HW} imply that, with high
probability, the following properties hold for every simplex $\sigma$
of the partition: 
(i) $\sigma$ is crossed by at most 
$O\left(\frac{m}{r}\log r\right)$ 
projected 2-planes;
(ii) the projected points that fall into $\sigma$ have at most
$s/r$ distinct $a_1$-coordinates and at most
$s/r$ distinct $b_1$-coordinates.  Further 
refining the simplices, if necessary, we can also assume that 
(iii) each simplex contains at most $n/r^3$ projected points.

Property (ii) is crucial. It asserts that the number of points 
of $S$ which induce the parabolas whose dual points project into a
fixed simplex is at most $2s/r$; more precisely, there are only
$s/r$ ``source'' points of $S$ and only
$s/r$ ``target'' points, so that each of these parabolas is of the
form $h^*_{a,b}$, where $a$ is one of the $s/r$ source points
and $b$ is one of the $s/r$ target points.
(Note, by the way, that the number of parabolas, $n/r^3$, in volved in
a subproblem is much smaller than the maximum possible value
$(s/r)^2$, when $r\gg 1$.)

We now apply Theorem~\ref{inc-gen} to each simplex $\sigma$; 
that is, to the set $C_\sigma$ of those parabolas whose (projected) 
dual points lie in $\sigma$, and to the set $P_\sigma$ of those
rotations whose (projected) dual 2-planes cross $\sigma$. 
Put $m_\sigma=|P_\sigma|$ and $n_\sigma=|C_\sigma|$.
We note that some rotations in $P_\sigma$ may be incident to 
no more than two parabolas in $C_\sigma$; these rotations
contribute $O(m_\sigma) = O\left(\frac{m}{r}\log r\right)$ 
to the overall incidence bound.
By Theorem~\ref{inc-gen} we thus have\footnote{%
  Here we cannot argue, as we did earlier, that the term 
  $m_\sigma$ is subsumed by the other terms, because of the
  possibility that some of the $m_\sigma$ rotations are incident 
  to only one or two parabolas in a subproblem.}
$$
I(P_\sigma,C_\sigma) = O\left(
m_\sigma^{1/3}n_\sigma +
m_\sigma^{2/3}n_\sigma^{1/3}(s/r)^{1/3} + m_\sigma \right) .
$$
Summing these bounds over all cells $\sigma$, we get an 
overall bound of
$$
\sum_\sigma I(P_\sigma,C_\sigma) = O^*\left(
r^3 \cdot \left( (m/r)^{1/3}n/r^3 +
(m/r)^{2/3}(n/r^3)^{1/3}(s/r)^{1/3} + m/r \right) \right) =
$$
$$
O^*\left( m^{1/3}n/r^{1/3} + rm^{2/3}n^{1/3}s^{1/3} + mr^2 \right) ,
$$
where, as above, $O^*(\cdot)$ hides polylogarithmic factors.

We also have to add to the bound incidences involving points, 
which are projections dual to parabolas, which lie on the 
boundaries of the cells of the cutting. 
Let $q=(a_1,a_2,b_1)$, the projection of a
(unique) point $\hat{h}_{a,b}$ be such a point.
Let $f$ denote the face whose relative interior contains 
$q$. If $f$ is a 2-face of some simplex $\sigma$, 
we can associate $q$ with $\sigma$: except for the 
single plane containing $f$, any other plane incident to 
$q$ must cross $\sigma$, and we can count the 
incidence within the subproblem of $\sigma$. The uncounted 
incidences, at most one per parabola, add up to at most $n$.

If $f$ is a vertex (so $q=f$) then any plane
through $f$ either bounds or crosses some adjacent simplex, 
so the total number of such incidences is
$O^*(r^3\cdot (m/r)) = O^*(mr^2)$. 

The harder situation is when $f$ is an edge. Again, if a plane
{\em crosses} $f$ at $q$, we can count this incidence within any 
adjacent simplex, arguing as in the case where $f$ is a 2-face. 
The difficult case is when the plane {\em contains} $f$, and we 
handle it as follows.

It is simpler to consider $f$ as a full line of intersection of
two sampled planes, rather than a single edge. (The decomposition,
though, has also other edges, obtained in the decomposition of
arrangement cells into simplices; these edges require a slightly
different treatment, given below.)
Let $q_1,\ldots,q_t$ be the projected dual points that lie on $f$,
and let $h^*_{a_i,b_i}$ denote the parabola corresponding to $q_i$,
for $i=1,\ldots,t$.
Consider the rotations $\tau$ whose dual 2-planes project to planes
containing $f$. Rotations $\tau$ of this kind which are incident to 
just one of the parabolas $h^*_{a_i,b_i}$ are easy to handle, because
the number of incidences involving these rotations is at most $m$ 
(for the fixed line $f$), for a total of $O^*(mr^2)$.

\begin{figure}[htbp]
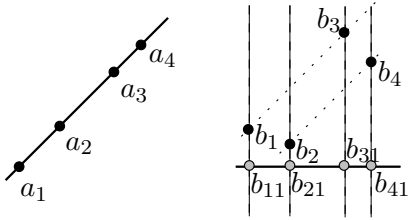

\begin{center}
\input abab.pstex_t
\caption{Many projected dual points lying on a common line: The situation in the
$xy$-plane.}
\label{abab}
\end{center}
\end{figure}

Consider then those rotations $\tau$ which are incident to at least
two of the parabolas $h^*_{a_i,b_i}$. Since the points
$(a_{i1},a_{i2},b_{i1})$ lie on a common line, it follows that the
points $a_i$ are also collinear in the original $xy$-plane, lying 
on a common line $\ell_0$. The points $b_i$ are not necessarily 
collinear, but they have the property that,
for any pair of indices $i\ne j$, the ratio 
$(b_{j1}-b_{i1})/(a_{j1}-a_{i1})$ is fixed.
See Figure~\ref{abab}.

Now if $\tau$ is incident to two parabolas
$h^*_{a_i,b_i}$, $h^*_{a_j,b_j}$, then $\tau$ maps $a_i$ to $b_i$
and $a_j$ to $b_j$. In particular, $|a_ia_j|=|b_ib_j|$. This, and the
fact that $(b_{j1}-b_{i1})/(a_{j1}-a_{i1})$ is fixed, imply that
$\tau$ maps $\ell_0$ to the line through $b_i$ and $b_j$, and that
the slope of this line has a {\em fixed absolute value} $\lambda$. 
Hence, considering, with no loss of generality, only lines of the
latter kind with positive slope, we can partition 
$\{q_1,\ldots,q_t\}$ into equivalence classes, so that, for each
class, all the corresponding points $b_i$ lie on a common line of
slope $\lambda$. Moreover, there is at most one rotation that is
incident to at least two parabolas from the same class (and no
rotation can be incident to two parabolas from different classes).
Thus the total number of incidences of this kind, for the fixed $f$,
is at most $t$. Summing over all lines $f$, we get a total of $O(n)$
such incidences.

In the preceding analysis we considered only intersection lines
between sampled planes, but, as noted, the cutting has additional
edges, interior to cells of the arrangement. We handle such edges in
almost the same way as above. That is, we consider such an edge $e$,
and argue, exactly as above, that the number of original incidences
involving points on $e$ and planes that contain $e$ is proportional to
the number $n_e$ of points on $e$ plus the number $m_e$ of planes
containing $e$. (Incidences involving planes that cross $e$ are also
handled exactly as above, wih the same resulting bound.)
The sum $\sum_e n_e$ is still at most $n$. For the
other sum $\sum_e m_e$, we note that the number of edges $e$ is
$O(r^3)$ (instead of $O(r^2)$ in the preceding analysis), but each
edge $e$ can be contained in at most $O\left(\frac{m}{r}\log r\right)$
planes, as follows easily from the $\eps$-net theory (this holds with
high probability, but we may assume that our sample does indeed have 
this property). Hence, we have
$\sum_e m_e = O^*(r^3\cdot(m/r)) = O^*(mr^2)$, the same bound as
above.

Altogether, the number of incidences is thus
$$
O^*\left( m^{1/3}n/r^{1/3}+mr^2+rm^{2/3}n^{1/3}s^{1/3}+n\right) .
$$
We now choose 
$$
r = \left( \frac{n^{2/3}}{m^{1/3}s^{1/3}} \right)^{3/4} =
\frac{n^{1/2}}{m^{1/4}s^{1/4}} .
$$
This choice of $r$ makes the first and third terms in the incidence
bound equal to each other, and they both dominate the second 
term, as is easily verified, using the fact that $n\le s^2$. 

Note also that $1\le r\le m$ when 
$$
\frac{n^{2/5}}{s^{1/5}} \le m\le \frac{n^{2}}{s} .
$$
Assume first that $m$ lies in this range. Then the incidence bound
becomes
$$
O\left(m^{5/12}n^{5/6}s^{1/12} + n\right) .
$$
When $m > n^{2}/s$, we use $r=1$ and get the bound 
$$
O\left( m^{1/3}n + m^{2/3}n^{1/3}s^{1/3} + m \right) .
$$
Since $n^{2}/s < m \le ns$, the second term dominates the two other
terms, and the bound is thus
$O\left( m^{2/3}n^{1/3}s^{1/3} \right)$.

Finally, when $m < n^{2/5}/s^{1/5}$, we use the Szemer\'edi-Trotter
bound in (\ref{weak23}), which is easily seen to yield the bound $O(n)$. 
Adding all these bounds, the theorem follows.
$\Box$

Using this bound, we can strengthen Corollary~\ref{mgek}, as follows.
\begin{corollary} \label{mgek1}
Let $C$ be a set of $n$ $h$-parabolas and $P$ a set of rotations, with
respect to a planar ground set $S$ of $s$ points. Then, for any
$k\ge 3$, the number $M_{\ge k}$ of rotations of $P$ incident to at
least $k$ parabolas of $C$ satisfies
$$
M_{\ge k} = O^*\left(
\frac{n^{10/7}s^{1/7}}{k^{12/7}} +
\frac{ns}{k^{3}} +
\frac{n}{k}
\right) .
$$
For $n=s^2$, the bound becomes
$$
M_{\ge k} = O^*\left( \frac{s^3}{k^{12/7}} \right) .
$$
\end{corollary}
\noindent{\bf Proof:}
The proof is similar to the proof of Corollary~\ref{mgek}, 
and we omit its routine details.
$\Box$

\section{Conclusion}
\label{sec:conc}

In this paper we have reduced the problem of obtaining a near-linear
lower bound for the number of distinct distances in the plane to a
problem involving incidences between points and a special class of
parabolas (or helices) in three dimensions. We have made significant 
progress in obtaining upper bounds for the number of such incidences, 
but we are still short of tightening these bounds to meet the 
conjectures on these bounds made in the introduction.

To see how far we still have to go, consider the bound in
Corollary~\ref{mgek1}, for the case $n=s^2$, which then becomes
$O^*(s^3/k^{12/7})$. (Here $M_{\ge k}$ coincides with $N_{\ge k}$ as
defined in (H3).) Moreover, we also have the Szemer\'edi-Trotter bound
$O(s^4/k^3)$, which is smaller than the previous bound for $k\ge s^{7/9}$.
Substituting these bounds in the analysis of (H3) and (H4), we get
$$
\frac{\left[s(s-1)-x\right]^2}{x} \le |K| =
N_{\ge 2} + \sum_{k\ge 3} (k-1) N_{\ge k} =
$$
$$
N_{\ge 2} +
O(s^3) \cdot \left[ 1 + \sum_{k=3}^{s^{7/9}} \frac{1}{k^{5/7}} +
  \sum_{k > s^{7/9}} \frac{s^4}{k^3} \right] = 
N_{\ge 2} + O(s^{29/9}) .
$$
It is fairly easy to show that $N_{\ge 2}$ is $O(s^{10/3})$, by
noting that $N_{\ge 2}$ can be upper bounded by 
$O\left(\sum_i |E_i|^2\right)$,
where $E_i$ is as defined in (H1). Using the upper bound
$|E_i|=O(s^{4/3})$ \cite{SST}, we get
$$
N_{\ge 2} = O\left(\sum_i |E_i|^2\right) = O(s^{4/3})\cdot
O\left(\sum_i |E_i|\right) = O(s^{10/3}) .
$$
Thus, at the moment, $N_{\ge 2}$ is the bottleneck in the above bound,
and we only get the (weak) lower bound
$\Omega(s^{2/3})$ on the number of distinct distances.
Showing that $N_{\ge 2}=O(s^{29/9})$ too (hopefully, a rather 
modest goal) would improve the lower bound to $\Omega(s^{7/9})$, 
still a rather weak lower bound.

Nevertheless, we feel that the reduction to incidences in three
dimensions is fruitful, because 

\noindent
(i) It sheds new light on the geometry of planar point sets, related to
the distinct distances problem.

\noindent
(ii) It gave us a new, and considerably more involved setup in which
the new algebraic technique of Guth and Katz could be applied. As
such, the analysis in this paper might prove useful for obtaining
improved incidence bounds for points and other classes of curves in
three dimensions. The case of points and circles is an immediate next
challenge.

Another comment is in order. Our work can be regarded as a special
variant of the complex version of the Szemer\'edi-Trotter theorem on
point-line incidences \cite{ST}. In the complex plane, the equation of
a line (in complex notation) is $w=pz+q$. Interpreting this equation
as a transformation of the real plane, we get a {\em homothetic map},
i.e., a rigid motion followed by a scaling. We can therefore rephrase
the complex version of the Szemer\'edi-Trotter theorem as follows.
We are given a set $P$ of $m$ pairs of points in the (real) plane, 
and a set $M$ of $n$ homothetic maps, and we seek an upper bound 
on the number of times a map $\tau\in M$ and a pair $(a,b)\in P$
``coincide'', in the sense that $\tau(a)=b$. In our work we only
consider ``complex lines'' whose ``slope'' $p$ has absolute value $1$
(these are our rotations), and the set $P$ is simply $S\times S$.

The main open problems raised by this work are:

\noindent
(a) Obtain a cubic upper bound for the number of rotations which map
only two points of the given ground planar set $S$ to another pair
of points of $S$. Any upper bound smaller than $O(s^{3.1358})$ would
already be a significant step towards improving the current lower bound of
$\Omega(s^{0.8641})$ on distinct distances \cite{KT}.

\noindent
(b) Improve further the upper bound on the number of incidences
between rotations and $h$-parabolas. 
Ideally, establish Conjectures 1 and 2.

\subsection*{Homage and Acknowledgments}

The bulk of the paper was written after the passing away of Gy\"orgy
Elekes in September 2008. However, the initial infrastructure, 
including the transformation of the distinct distances problem 
to an incidence problem in three dimensions, and many other steps, 
is due to him.  As a matter of fact, it was already discovered 
by Elekes about 10 years ago, and lay dormant since then, mainly
because of the lack of effective tools for tackling the incidence problem. 
These tools became available with the breakthrough result of Guth and
Katz~\cite{GK} in December 2008, and have made this paper possible.
Thanks are due to M\'arton Elekes, who was a driving force in
restarting the research on this problem.

Many thanks are due to Haim Kaplan,
for many hours of helpful discussions concerning
the work in this paper. As mentioned, the construction in
Lemma~\ref{lem:lower} is due to him.

Finally, thanks are also due to
Jozsef Solymosi for some helpful comments on the technique
used in the paper.

\end{document}